\documentclass[10pt,journal,compsoc]{IEEEtran}
\ifCLASSOPTIONcompsoc
  \usepackage[nocompress]{cite}
\else
  \usepackage{cite}
\fi
\ifCLASSINFOpdf
\else
\fi
\usepackage[utf8]{inputenc}
\usepackage{amsmath,amssymb,amsfonts}
\usepackage{cite}

\usepackage{amsmath,amssymb,amsfonts}
\usepackage{cite}
\usepackage{multicol}
\usepackage{siunitx}
\usepackage{tikz}
\usepackage[T1]{fontenc}
\usepackage{algorithmic}
\usepackage{stfloats}
\usepackage{graphicx}
\usepackage{textcomp}
\usepackage{xcolor}
\usepackage[justification=centering]{caption}
\usepackage{pifont}
\usepackage{longtable}
\usepackage{amsthm}

\usepackage{hyperref}
\usepackage{mathrsfs}
\usepackage{booktabs}
\usepackage{hyperref}
\usepackage{cases}
\usepackage{comment}
\usepackage{empheq} 
\usepackage{lipsum,booktabs}
\usepackage{caption}
\usepackage{subcaption}
\allowdisplaybreaks[4]
\usepackage{url}  
\usepackage{graphicx}  
\usepackage[ruled,vlined,linesnumbered]{algorithm2e}
\usepackage{floatpag} 
\usepackage{float}
\usepackage{tabularx}
\usepackage{booktabs}
\usepackage[labelfont=bf,format=plain,justification=raggedright,singlelinecheck=false]{caption}

\usepackage{url}  
\usepackage{graphicx}  
\usepackage[ruled,vlined,linesnumbered]{algorithm2e}

\newcommand*\circled[1]{\tikz[baseline=(char.base)]{
            \node[shape=circle,draw,inner sep=2pt] (char) {#1};}}

\usepackage{listings}
\def\BibTeX{{\rm B\kern-.05em{\sc i\kern-.025em b}\kern-.08em
    T\kern-.1667em\lower.7ex\hbox{E}\kern-.125emX}}
\begin{document}
\captionsetup[figure]{labelfont={rm},labelformat={default},labelsep=period,name={Fig.}}
%
\title{Power Source Allocation for 
RIS-aided Integrating Sensing, Communication, and Power Transfer Systems Based on NOMA}
%
%
%
%

\author{Yue Xiu,~Yang Zhao,~Chenfei Xie,~Fatma Benkhelifa~\IEEEmembership{Member,~IEEE}, ~Songjie Yang,~Wanting~Lyu,~Chadi Assi~\IEEEmembership{Fellow,~IEEE},~Ning Wei~\IEEEmembership{Member,~IEEE}
~ \\
\thanks{The corresponding author is Ning Wei.}
}

\IEEEtitleabstractindextext{%
\begin{abstract}
This paper proposes a novel communication system framework based on a reconfigurable intelligent surface (RIS)-aided integrated sensing, communication, and power transmission (ISCPT) communication system. RIS is used to improve transmission efficiency and sensing accuracy. In addition, non-orthogonal multiple access (NOMA) technology is incorporated in RIS-aided ISCPT systems to boost the spectrum utilization efficiency of RIS-aided ISCPT systems. We consider the power minimization problem of the RIS-aided ISCPT-NOMA system. Power minimization is achieved by jointly optimizing the RIS phase shift,  decoding order, power splitting (PS) factor, and transmit beamforming while satisfying quality of service (QoS), radar target sensing accuracy, and energy harvesting constraints. Since the objective function and constraints in the optimization problem are non-convex, the problem is an NP-hard problem. To solve the non-convex problem, this paper proposes a block coordinate descent (BCD) algorithm. Specifically, the non-convex problem is divided into four sub-problems: i.e. the transmit beamforming,  RIS phase shift, decoding order and PS factor optimization subproblems. We employ semidefinite relaxation (SDR) and successive convex approximation (SCA) techniques to address the transmit beamforming optimization sub-problem. Subsequently, we leverage the alternating direction method of multipliers (ADMM) algorithm to solve the RIS phase shift optimization problem. As for the decoding order optimization, we provide a closed-form expression. For the PS factor optimization problem, the SCA algorithm is proposed. Simulation results illustrate the effectiveness of our proposed algorithm and highlight the balanced performance achieved across sensing, communication, and power transfer.
\end{abstract}

}

\maketitle

\IEEEdisplaynontitleabstractindextext

%
\IEEEpeerreviewmaketitle

\IEEEraisesectionheading{\section{Introduction}\label{sec:introduction}}
\IEEEPARstart{W}{ireless} communication and Internet-of-Things (IoT) technology have recently developed rapidly. The number of sensing and wireless communication devices has exponentially increased\cite{b1,b2,b3}. Consequently, in the sixth-generation (6G) mobile communication system, to fulfill the demands of sensing, energy harvesting, and massive access, the integration of sensing, communication, and power transfer (ISCPT) technology, along with non-orthogonal multiple access (NOMA) technologies, has attracted significant discussion within academia and industry\cite{b4,b5,b6,b7}. Compared to traditional communication systems, ISCPT signifies a progression in wireless technology, leveraging similar hardware platforms and signal processing algorithms for wireless communication, energy harvesting, and radar sensing\cite{b8}. However, in the ISCPT communication system\cite{b9,b10}, when the channels of massive IoT devices are highly correlated, or the system is overloaded due to the limited spatial degree of freedom (DoF), the communication users will suffer from severe inter-user interference. To overcome this limitation, the emergence of NOMA technology provides technical solutions to address these challenges. Unlike traditional Orthogonal Multiple Access (OMA)\cite{b9}, NOMA enables many devices to share the same resources, meeting the communication needs of many devices and enhancing spectrum efficiency\cite{b11,b12}. Therefore, tackling the challenges the ISCPT-NOMA system framework poses has become a pressing concern. However,  the existing literature primarily focuses on integrating sensing and communication (ISAC) into the NOMA system\cite{b13,b14,b15,b16,b17,b18}, and there is currently no literature discussing the ISCPT-NOMA system.

 Specifically, the authors proposed an ISAC scheme based on multi-domain NOMA of\cite{b13}. The proposed ISAC-NOMA scheme can transmit non-orthogonal data streams in TF (time-frequency) and DD (delay-Doppler) domains while detecting the target. In\cite{b14}, a receiver structure combining zero-forcing (ZF) and maximum likelihood (ML) elements was proposed to eliminate the mutual interference between communication and radar systems and improve the transmission rate of ISAC-NOMA systems. In\cite{b15}, the authors presented a semi-integrated sensing and communication (Semi-ISAC) concept and studied the channel capacity of the Semi-ISAC-NOMA system. Compared with the traditional ISAC-OMA scheme, the transmission rate was further improved. In\cite{b16}, the authors presented a novel iterative optimization method to provide spectral efficiency and improve the transmission rate of the ISAC-NOMA system while maintaining sensing and communication performance. In\cite{b17}, the authors proposed an efficient dual decomposition penalty-based algorithm for the ISAC-NOMA system with high correlation channels. The simulation results showed that the transmission rate under the limitation of sensing accuracy was lower than that of the traditional ISAC system. In\cite{b18}, the authors proposed a joint communication, sensing, and multi-layer computing framework for ISAC-NOMA systems. In this framework, a multi-functional BS simultaneously performed target sensing and provided edge computing services to nearby users. While the studies mentioned above have made significant contributions to the field of ISAC-NOMA, they have not considered applying the NOMA framework within the integrated sensing, communication, and power transmission (ISCPT) system. 
 This paper aims to bridge that gap by introducing the NOMA framework into the ISCPT system, opening up new possibilities and potential improvements in this emerging area.

Furthermore, due to the increased complexity of 6G communication scenarios, transmission signals are easily blocked. The Reconfigurable Intelligent Surface (RIS) offers a promising and cost-effective solution to manipulate and adjust the wireless communication environment, effectively addressing transmission signal blockages\cite{b19}. As a result, both the RIS-assisted ISAC-NOMA system and RIS-assisted simultaneous wireless information and power transfer (SWIPT) system have garnered widespread attention from academia and industry\cite{b20,b21,b22,b23,b24,b25,b26}. In \cite{b20}, the authors explored the potential of RIS to enhance radar sensing in the NOMA-ISAC system. They jointly optimized active beamforming, power sharing coefficients, and passive beamforming to maximize the radar beam pattern gain. In \cite{b21}, the authors considered an RIS-assisted ISAC-NOMA multicast-unicast integrated communication system, where multicast signals were used for sensing and communication, and unicast signals were used for wireless communication to improve the transmission performance of the ISAC-NOMA system. They proposed an alternating optimization algorithm based on semidefinite relaxation (SDR) and successive rank-constrained relaxation (SROCR) to solve the problem. In \cite{b22}, an RIS-assisted ISAC-NOMA system was proposed to provide a virtual line-of-sight (LoS) link to radar targets, addressing significant path loss or blockage problems. Specifically, the RIS beam pattern gain was derived and used as a sensing metric. In \cite{b23}, the authors studied an RIS-assisted ISAC-NOMA system, formulating a max-min problem to optimize the sensing beam pattern under communication rate constraints by jointly optimizing power allocation, active beamforming, and the RIS phase profile. Their algorithm effectively improved the minimum beam pattern gain (MBPG) compared to conventional baselines. In \cite{b24}, a novel RIS-enhanced SWIPT system based on an electromagnetic compatibility framework was proposed. In \cite{b25}, the authors investigated RIS-assisted SWIPT networks with rate-splitting multiple access (RSMA), maximizing energy efficiency (EE) by optimizing beamforming vectors, power splitting (PS) ratio, and the discrete-valued phase shifts of the RIS. In \cite{b26}, a SWIPT system based on UAV-RIS was studied, where a robust algorithm based on deep reinforcement learning (DRL) was developed to improve the communication quality of service (QoS) in dynamic wireless environments.
In \cite{b27}, the authors examined the fairness of simultaneously transmitting and reflecting (STAR-RIS) aided ISAC-NOMA systems. They maximized fairness between communication users and sensing targets by jointly optimizing the transmit beamforming vector of the BS and the coefficient matrix of STAR-RIS. It is apparent that the literature mentioned above considers RIS-assisted ISAC or SWIPT systems separately. To the best of the authors' knowledge, no literature has yet considered the RIS-assisted ISCPT system.

Based on the above discussion, although a lot of research has been conducted on ISAC-NOMA, ISAC-SWIPT, and RIS-aided SWIPT systems, the future 6G communication environment will be more complex and the demand for sensing services and energy harvesting in IoT networks will also increase greatly. This situation inspires us to integrate ISAC and SWIPT to meet users' performance requirements in IoT networks.  However, current research on the RIS-aided ISCPT-NOMA system is insufficient. This paper studies the power minimization problem of the RIS-aided ISCPT-NOMA system by optimizing the successive interference cancellation (SIC), the transmit beam, the receive PS ratio, and the RIS phase shift under the constraints of user QoS requirements, target detection accuracy, energy harvesting, and RIS unit modulus reflection coefficients to minimize the BS transmit power. Due to the more complex SIC caused by wireless channel changes and the nature of the considered formulation, the resulting optimization problem is nondeterministic polynomial (NP)-hard. Thus, we need to design a practical algorithm for the power source allocation problem of the RIS-aided ISCPT-NOMA system. The contributions of this paper are summarized as follows:
\begin{itemize}
\item This paper investigates a novel RIS-aided ISCPT-NOMA system. The same resource block serves multiple users simultaneously, and the signal transmitted can also detect point targets while sending information to users. Additionally, users can collect energy. By being subject to point target sensing accuracy limitations and SINR constraints, the RIS can enhance the energy efficiency of the ISCPT-NOMA system.  Therefore, the transmit power minimization problem of RIS-aided ISCPT-NOMA systems is formulated as a joint optimization problem of SIC decoding order, PS ratio, RIS phase shift, and transmit beamforming. 
\item To solve the optimization problem, we divide the problem into four sub-problems. Specifically, a transmit beamforming optimization algorithm based on successive convex approximation (SCA) is proposed for the first sub-problem. We offer an RIS phase shift optimization algorithm based on the alternating direction method of multipliers (ADMM) for the second sub-problem. A SIC decoding sequence design algorithm based on SCA is proposed for the third sub-problem. Similarly, the PS optimization problem can be solved using SCA. Finally, the proposed block coordinate descent (BCD) algorithm, which consists of four sub-algorithms, is iterated until convergence.
\item By using MATLAB simulations, we show the performance of the proposed BCD algorithm for the transmit beamforming vector, RIS phase shift, SIC decoding order, and PS, in comparison to baseline algorithms. Specifically, we show that the considered approach can significantly reduce the system's transmit power requirements. The BS transmit power of the RIS-assisted ISCPT-NOMA system is significantly lower than that of the network without RIS assistance. Furthermore, it is evident that as the QoS requirements become more stringent, the target positioning accuracy of RIS-assisted ISCPT-NOMA decreases.
\end{itemize}

\textbf{Organization:} This paper comprises the following sections. Section \ref{II} introduces the novel RIS-aided ISCPT-NOMA system model. Section \ref{III} presents the problem formulation. Section \ref{IV} analyzes the joint transmit beamforming design, RIS reflection beamforming design, PS parameters, the SIC decoding sequence optimization algorithm, and the convergence of the proposed algorithm. Section \ref{V} demonstrates the performance of the proposed algorithm through numerical results. Finally, Section \ref{VI} summarizes the paper.

\section{System Model and Problem Formulation}\label{II}
\begin{figure}[htbp]
  \centering
  \includegraphics[scale=0.5]{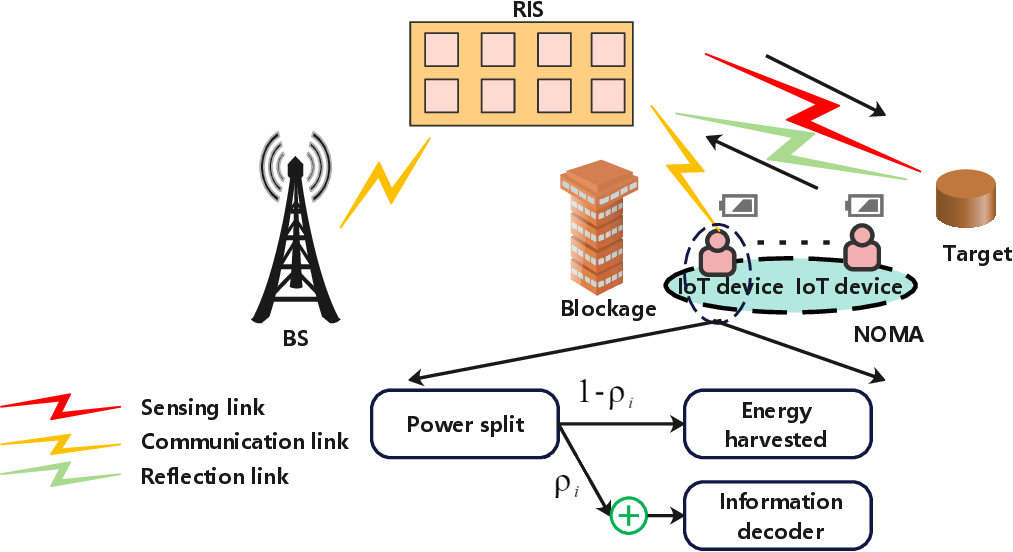}
  \captionsetup{justification=centering}
  \caption{Illustration of the considered RIS-aided ISCPT-NOMA system.}
\label{FIGURE1}
\end{figure}
As shown in Fig.~\ref{FIGURE1}, we consider an RIS-aided ISCPT-NOMA system, where a BS is equipped with $N_{T}$ transmit antennas and $N_{R}$ receive antennas. In this RIS-aided ISCPT-NOMA system, the number of single-antenna IoT devices is $K$, where $K<N_{T}$ and every receiver can harvest energy. The IoT devices operate as shown in Fig.~\ref{FIGURE1}. Without loss of generality, we assume that the locations of the IoT devices are random. Additionally, to reduce the signal loss caused by obstacles, an RIS with $M$ reflection elements is deployed in the RIS-aided ISCPT-NOMA system. Subsequently, we introduce the communication signal model and sensing signal model in detail.

The received signal from the BS at the $k$-th IoT device is expressed as
\begin{small}
\begin{align}
&y_{k}[n]=\sqrt{\rho_{k}}\left(\mathbf{h}^{H}_{u,k}\boldsymbol{\Theta}\mathbf{G}\mathbf{x}_{k}[n]+\sum\nolimits_{k^{\prime}\neq k,k^{\prime}=1}^{K}\mathbf{h}^{H}_{u,k^{\prime}}\boldsymbol{\Theta}\mathbf{G}\mathbf{x}_{k}[n]\right.\nonumber\\
&\left.+n_{k}[n]\right)+\bar{n}_{k}[n],k,k^{\prime}\in\{1,\ldots,K\}, \label{pro1}
\end{align}
\end{small}%
where $\rho_{k}$ $(0\leq\rho_{k}\leq1)$ is a portion of signal power allocated to the IoT devices, $\mathbf{x}_{k}=\mathbf{w}_{k}s_{k}[n]\in\mathbb{C}^{N_{T}\times 1}$ is the transmitted signal, $s_{k}[n]\in\mathbb{C}^{1\times 1}$ is the data stream of the $n$-th time, and $\mathbb{E}\{s_{k}[n]s_{k}^{H}[n]\}=1$, $\mathbf{w}_{k}\in\mathbb{C}^{N_{T}\times 1}$ is the beamforming vector. $n_{k}[n]\sim\mathcal{CN}(0,\sigma_{k}^{2})$ and $\bar{n}_{k}[n]\sim\mathcal{CN}(0,\bar{\sigma}_{k}^{2})$ are the additive white Gaussian noise (AWGN) at the IoT device. $\mathbf{G}\in\mathbb{C}^{M\times N_{T}}$ and $\mathbf{h}_{u,k}\in\mathbb{C}^{M\times 1}$ are the channels from the BS to the RIS and the channels from the RIS to the $k$-th IoT device, respectively.
According to\cite{b28,b29,b30}, $\mathbf{G}$ is formulated as 
\begin{small}
\begin{align}
\mathbf{G}=\alpha_{G}\sqrt{\beta/(\beta+1)}\mathbf{G}^{LoS}+\alpha_{G}\sqrt{1/(\beta+1)}\mathbf{G}^{NLoS},\label{pro2}
\end{align}
\end{small}%
where $\alpha_{G}$ is path loss depending on distance, and $\beta$ is the Rician factor. $\mathbf{G}^{LoS}\in\mathbb{C}^{M\times N_{T}}$ and $\mathbf{G}^{NLoS}\in\mathbb{C}^{M\times N_{T}}$ are line-of-sight (LoS) channel components and non-line-of-sight (NLoS) channel components, where $\mathbf{G}^{LoS}=\mathbf{a}_{M}(\theta_{DOA})\mathbf{a}_{N_{T}}^{T}(\theta_{DoD})$, $\mathbf{a}_{N_{T}}(\theta_{DoD})=\left[1,e^{-j\pi\sin\theta_{DoD}},\ldots,e^{-j(N_{T}-1)\pi\sin\theta_{DoD}}\right]^{T}\in\mathbb{C}^{N_{T}\times 1}$ and $\mathbf{a}_{M}(\theta_{DoA})=\left[1,\right.$
$\left.e^{-j\pi\sin\theta_{DoA}},\ldots,\right.$
$\left.e^{-j(M-1)\pi\sin\theta_{DoA}}\right]^{T}\in\mathbb{C}^{M\times 1}$ are array response vectors. $\theta_{DoD}$ and $\theta_{DoD}$ are the direction of departure (DoD) and direction of arrival (DoA).  Similarly, channel $\mathbf{h}_{u,k}$ is denoted as
\begin{small}
\begin{align}
\mathbf{h}_{u,k}=\alpha_{u}\sqrt{\beta_{u}/(\beta_{u}+1)}\mathbf{h}_{u,k}^{LoS}+\alpha_{u}\sqrt{1/(\beta_{u}+1)}\mathbf{h}_{u,k}^{NLoS},\label{pro3}
\end{align}
\end{small}%
where $\alpha_{u}$ is the path loss dependent on distance,  and $\beta_{u}$ is the Rician factor. $\mathbf{h}_{u,k}^{LoS}\in\mathbb{C}^{M\times 1}$ and $\mathbf{h}_{u,k}^{NLoS}\in\mathbb{C}^{M\times 1}$ are the LoS channel components and the NLoS channel components, where $\mathbf{h}_{u,k}^{LoS}=\mathbf{a}_{M}(\theta_{DoD})$ and $\mathbf{a}_{M}(\theta_{DoD})=[1,e^{-j\pi\sin\theta_{DoD}},\ldots,e^{-j(N-1)\pi\sin\theta_{DoD}}]^{T}\in\mathbb{C}^{M\times 1}$ is array response vector. In addition, $\mathbf{h}_{u,k}^{NLoS}$ is the NLoS component, where each element of $\mathbf{h}_{u,k}^{NLoS}$ is i.i.d. complex Gaussian distributed with zero mean and unit variance\cite{b39}. $\boldsymbol{\Theta}\in\mathbb{C}^{M\times M}$ denotes the RIS reflection matrix and $\boldsymbol{\Theta}=\mathrm{diag}\{\boldsymbol{\theta}\}$, where $\boldsymbol{\theta}=[\theta_{1},\ldots,\theta_{M}]^{T}\in\mathbb{C}^{M\times 1}$, $\theta_{m}=e^{j\nu_{m}}$, $0\leq \nu_{m}\leq 2\pi$, $m\in\{1,\ldots,M\}$, and the RIS reflection coefficients satisfy $|\theta_{m}|=1$.
The transmit power constraint is expressed as \begin{small}
\begin{align}
\mathrm{tr}\left(\sum\nolimits_{k=1}^{K}\mathbf{w}_{k}\mathbf{w}_{k}^{H}\right)\leq P_{T},\label{pro4}
\end{align}
\end{small}%
where $P_{T}$ is the total transmit power constraint, and $\boldsymbol{w}_{k}$ is the beamforming vector from the BS.

\subsection{Radar Sensing Signal Model}
The transmitted signal reaches the target via a reflection link and then is reflected to the BS through the same link. As a result, the baseband echo signal, which is reflected by the target and captured by the BS, can be described as follows
\begin{align}
\mathbf{y}_{s}[n]=\sum\nolimits_{k=1}^{K}\alpha\mathbf{h}_{s}(\boldsymbol{\Theta})\mathbf{w}_{k}s_{k}[n]+\mathbf{n}_{s},\label{pro5}
\end{align}
where $\mathbf{h}_{s}(\boldsymbol{\Theta})=(\mathbf{h}_{s}^{H}\boldsymbol{\Theta}\mathbf{G})^{H}\mathbf{h}_{s}^{H}\boldsymbol{\Theta}\mathbf{G}\in\mathbb{C}^{N_{T}\times N_{T}}$, $\alpha$ is the radar cross section (RCS), $\mathbf{h}_{s}\in\mathbb{C}^{M\times 1}$ is the channel from the RIS to the target, and $\mathbf{n}_{s}\in\mathbb{C}^{N_{T}\times 1}\sim\mathcal{CN}(\boldsymbol{0},\sigma_{s}^{2}\boldsymbol{I})$ is the AWGN. According to \cite{b31}, without loss of generality, we assume that the RIS target link is LoS. Specifically, $\mathbf{h}_{s}=\alpha_{s}\mathbf{a}_{M}(\varphi_{DoA})$, where $\varphi_{DoA}$ represents the DoA of RIS and $\mathbf{a}_{M}(\varphi_{DoA})=\left[1,e^{-j\pi\sin\varphi_{DoA}},\ldots,e^{-j(M-1)\pi\sin\varphi_{DoA}}\right]^{T}\in\mathbb{C}^{M\times 1}$, $\alpha_{s}$ is channel gain. For radar, it is assumed that the current angle/range cell we are detecting contains the target. The known DoA is used to calculate the target detection probability or estimate the Cramer-Rao bound (CRB). Furthermore, due to the semi-static nature of the BS-RIS channel, $\mathbf{G}$ can be estimated using a two-time scale channel estimation framework\cite{b32}. According to\cite{b8}, radar sensing is analyzed by collecting $N$ samples. Therefore, we need to combine $N$ samples, and the radar sensing signal is given by (\ref{pro6}) at the top of this page in formulation (\ref{pro6}),  
$\mathbf{S}=[\mathbf{s}[1],\ldots,$ $\mathbf{s}[N]]\in\mathbb{C}^{N_{T}\times N}$, and $\mathbf{N}_{s}=[\mathbf{n}_{s}[1],\ldots,\mathbf{n}_{s}[N]]\in\mathbb{C}^{N_{T}\times N}$. It is noted that the equivalent channel $\mathbf{h}_{s}(\boldsymbol{\Theta})$ is known at the BS based on the above assumptions.
\begin{figure*}[ht] 
\centering 
\vspace*{6pt} 
\begin{small}
\begin{align}
&\mathbf{Y}_{s}=\left[\begin{matrix}
\mathbf{y}_{s}[1]\\
\vdots\\
\mathbf{y}_{s}[N]
  \end{matrix}
  \right]=\left[\begin{matrix}
\sum_{k=1}^{K}\alpha(\mathbf{h}_{s}^{H}\boldsymbol{\Theta}\mathbf{G})^{H}\mathbf{h}_{s}^{H}\boldsymbol{\Theta}\mathbf{G}\mathbf{w}_{k}s_{k}[1]+\mathbf{n}_{s}[1]\\
\vdots\\
\sum_{k=1}^{K}\alpha(\mathbf{h}_{s}^{H}\boldsymbol{\Theta}\mathbf{G})^{H}\mathbf{h}_{s}^{H}\boldsymbol{\Theta}\mathbf{G}\mathbf{w}_{k}s_{k}[N]+\mathbf{n}_{s}[N]
  \end{matrix}
  \right]=\left[\begin{matrix}
\sum_{k=1}^{K}\alpha(\mathbf{h}_{s}^{H}\boldsymbol{\Theta}\mathbf{G})^{H}\mathbf{h}_{s}^{H}\boldsymbol{\Theta}\mathbf{G}\mathbf{w}_{k}s_{k}[1]\\
\vdots\\
\sum_{k=1}^{K}\alpha(\mathbf{h}_{s}^{H}\boldsymbol{\Theta}\mathbf{G})^{H}\mathbf{h}_{s}^{H}\boldsymbol{\Theta}\mathbf{G}\mathbf{w}_{k}s_{k}[N]
  \end{matrix}
  \right]+\left[\begin{matrix}
\mathbf{n}_{s}[1]\\
\vdots\\
\mathbf{n}_{s}[N]
  \end{matrix}
  \right]=\alpha\mathbf{h}_{s}(\boldsymbol{\Theta})\mathbf{W}\nonumber\\
  &\mathbf{S}+\mathbf{N}_{s},\label{pro6}
\end{align}
\end{small}
\hrulefill
\end{figure*}

\begin{table}[!ht]
\centering
\caption{Notations.}
\label{notations}
\begin{tabular}{cp{5.0cm}}
\toprule[1pt]
\textbf{Symbol} & \textbf{Descriptions} \\
\midrule
$K$ & The number of UEs. \\
$N_{T}/N_{R}$ &  The number of antennas. \\
$M$ & The number of RIS reflection elements. \\
$N$ & The number of sensing signal samples. \\
$\rho_{k}$ & The signal power allocation factor. \\
$y_{k}$ & Received signal. \\
$n_{k}/\bar{n}_{k}$ & AWGN. \\
$\boldsymbol{\Theta}$ & The phase shift matrix of RIS. \\
$\boldsymbol{x}_{k}$ & The transmitted signal. \\
$\boldsymbol{w}_{k}$ & The transmitted beamforming. \\
$\boldsymbol{s}_{k}$ & The data stream. \\
$\alpha_{G}/\alpha_{u}$ & The path loss factor. \\
$\beta/\beta_{u}$ & The Rician factor. \\
$\boldsymbol{G}^{LoS}/\boldsymbol{h}_{u,i}^{LoS}$ & The Los channel. \\
$\boldsymbol{G}^{NLoS}/\boldsymbol{h}_{u,i}^{NLoS}$ & The NLoS channel. \\
$\theta_{DoA}/\theta_{DoD}$ & The AoA/DoA. \\
$P_{T}$ & The total transmit power. \\
\bottomrule[1pt]
\end{tabular}\label{TA1}
\end{table}

\section{Problem Formulation}\label{III}
In this section, we calculate the performance metrics, including the SINR for communication and the CRB for radar sensing. These metrics are essential for constructing the system's optimization problem. Subsequently, a power minimization problem is formulated while considering constraints on the communication SINR and radar sensing accuracy CRB to reduce power usage.

\subsection{Communication System SINR and Energy Harvesting (EH)}
In contrast to traditional communication systems, we focus on the NOMA communication system. Considering that the equivalent channel gain determines the SIC decoding order of the NOMA system. The introduction of RIS assistance in the NOMA system increases the complexity of the channel because the channel gain may fluctuate due to variations in the RIS phase shift matrix. Represented by $d(k)$, the decoding order of the $k$-th IoT device indicates that $d(k)=i$ denotes the $k$-th IoT device as the $i$-th signal to be decoded at the receiver. As a result, the SINR of the $k$-th IoT device can be expressed as 
\begin{small}
\begin{align}
\mathrm{SINR}_{k}=\frac{\rho_{k}|\mathbf{h}^{H}_{u,k}\boldsymbol{\Theta}\mathbf{G}\mathbf{w}_{k}|^{2}}{\rho_{k}\sum_{d(j)> d(k)}|\mathbf{h}^{H}_{u,k}\boldsymbol{\Theta}\mathbf{G}\mathbf{w}_{j}|^{2}+\rho_{k}\sigma_{k}^{2}+\bar{\sigma}_{k}^{2}},\label{pro7}
\end{align}
\end{small}%
Then, $d(k)\leq d(\bar{k})$ is assumed\cite{b33}, 
and the SINR of the $\bar{k}$-th IoT device decoding information signal of the $k$-th IoT device can be given by
\begin{small}
\begin{align}
\mathrm{SINR}_{\bar{k}-k}=\frac{\rho_{\bar{k}}|\mathbf{h}^{H}_{u,\bar{k}}\boldsymbol{\Theta}\mathbf{G}\mathbf{w}_{\bar{k}}|^{2}}{\rho_{\bar{k}}\sum_{d(j)> d(k)}|\mathbf{h}^{H}_{u,\bar{k}}\boldsymbol{\Theta}\mathbf{G}\mathbf{w}_{j}|^{2}+\rho_{\bar{k}}\sigma_{\bar{k}}^{2}+\bar{\sigma}_{\bar{k}}^{2}}.\label{pro8}
\end{align}
\end{small}%
To ensure that the $\bar{k}$-th IoT device can decode the information of the $k$-th IoT device with the decoding order $d(k)\leq d(\bar{k})$, the SIC decoding condition $\mathrm{SINR}_{k}\leq\mathrm{SINR}_{\bar{k}-k}$ should be satisfied\cite{b34}. For example, assuming the decoding order for the three IoT devices is $d(k)=k, \text{where}~k=1,2,3$, the SIC decoding conditions for IoT device $2$ and IoT device $3$ should satisfy the following conditions: $\mathrm{SINR}_{2-1}\geq\mathrm{SINR}_{1}$, $\mathrm{SINR}_{3-1}\geq\mathrm{SINR}_{1}$, and $\mathrm{SINR}_{3-2}\geq\mathrm{SINR}_{2}$. In addition, the energy harvested signal for the $k$-th IoT device can be expressed as 
\begin{small}
\begin{align}
y_{k}^{EH}=\sqrt{1-\rho_{k}}\left(\sum\nolimits_{k=1}^{K}\mathbf{h}^{H}_{u,k}\boldsymbol{\Theta}\mathbf{G}\mathbf{x}_{k}+n_{k}\right).\label{pro9}   
\end{align}
\end{small}%
Then, the harvested power of $k$-th IoT device can
be written as
\begin{small}
\begin{align}
E_{k}=\eta_{k}(1-\rho_{k})\left(\sum\nolimits_{k=1}^{K}|\mathbf{h}^{H}_{u,k}\boldsymbol{\Theta}\mathbf{G}\mathbf{w}_{k}|^{2}+\sigma_{k}^{2}\right),\label{pro10}   
\end{align}
\end{small}%
in which the power transform efficiency at the IoT device $k$ is $\eta_{k}\in (0, 1]$, with a focus on normalized time. Consequently, the harvested power is equivalently referred to as the harvested energy.

\subsection{Radar Sensing CRB}
Parameter estimation is crucial in radar sensing, with the accuracy typically evaluated using the CRB, which serves as a minimum threshold for unbiased estimators. In our specific scenario, the focus is on estimating the DoA denoted by $\theta=\varphi_{DoA}$. The CRB for estimating $\theta$ is derived by vectorizing the received signal $\mathbf{Y}_{s}$
\begin{small}
\begin{align}
\mathbf{y}_{s}=\mathrm{vec}\left(\alpha\mathbf{h}_{s}(\boldsymbol{\Theta})\mathbf{W}\mathbf{S}+\mathbf{N}_{s}\right)=\alpha\mathrm{vec}\left(\mathbf{h}_{s}(\boldsymbol{\Theta})\mathbf{W}\mathbf{S}\right)+\mathbf{n}_{s},\label{pro11}
\end{align}
\end{small}%
where $\mathbf{n}_{s}=\mathrm{vec}\{\mathbf{N}_{s}\}\in\mathbb{C}^{N_{T}N\times 1}$. We let $\boldsymbol{\tau}=[\theta,\boldsymbol{\alpha}^{T}]^{T}\in\mathbb{C}^{3\times 1}$, and $\boldsymbol{\tau}$ is the three-tuple number. $\boldsymbol{\alpha}=[\mathrm{Re}\{\alpha\},\mathrm{Im}\{\alpha\}]$, $\boldsymbol{\kappa}=\alpha\mathrm{vec}((\mathbf{h}_{s}^{H}\boldsymbol{\Theta}\mathbf{G})^{H}\mathbf{h}_{s}^{H}\boldsymbol{\Theta}\mathbf{G}\mathbf{W}\mathbf{s})\in\mathbb{C}^{N_{T}N\times 1}$ is the noise-free echo signal, and the Fisher information matrix (FIM) and CRB matrix are used for estimating $\boldsymbol{\tau}$ as $\mathbf{FIM}\in\mathbb{C}^{3\times 3}$ and $\mathbf{C}\in\mathbb{C}^{3\times 3}$, respectively. According to\cite{b30}, the observation signals $\mathbf{y}_{s}\sim\mathcal{CN}(\boldsymbol{\eta},\mathbf{R}_{s})$ and $\mathbf{R}_{s}=\sigma^{2}_{s}\mathbf{I}$; therefore, Fisher's information matrix can be expressed as
\begin{small}
\begin{align}
&\mathbf{FIM}=\frac{2}{\sigma_{s}^{2}}\nonumber\\
&\left[\begin{matrix}
\mathrm{Re}\left\{\frac{\partial^{H}\boldsymbol{\kappa}}{\partial \tau_{1}}\frac{\partial\boldsymbol{\kappa}}{\partial \tau_{1}}\right\}&\mathrm{Re}\left\{\frac{\partial^{H}\boldsymbol{\kappa}}{\partial \tau_{1}}\frac{\partial\boldsymbol{\kappa}}{\partial \tau_{2}}\right\}&\mathrm{Re}\left\{\frac{\partial^{H}\boldsymbol{\kappa}}{\partial \tau_{1}}\frac{\partial\boldsymbol{\kappa}}{\partial \tau_{3}}\right\}\\
\mathrm{Re}\left\{\frac{\partial^{H}\boldsymbol{\kappa}}{\partial \tau_{2}}\frac{\partial\boldsymbol{\kappa}}{\partial \tau_{1}}\right\}&\mathrm{Re}\left\{\frac{\partial^{H}\boldsymbol{\kappa}}{\partial \tau_{2}}\frac{\partial\boldsymbol{\kappa}}{\partial \tau_{2}}\right\}&\mathrm{Re}\left\{\frac{\partial^{H}\boldsymbol{\kappa}}{\partial \tau_{2}}\frac{\partial\boldsymbol{\kappa}}{\partial \tau_{3}}\right\}\\
\mathrm{Re}\left\{\frac{\partial^{H}\boldsymbol{\kappa}}{\partial \tau_{3}}\frac{\partial\boldsymbol{\kappa}}{\partial \tau_{1}}\right\}&\mathrm{Re}\left\{\frac{\partial^{H}\boldsymbol{\kappa}}{\partial \tau_{3}}\frac{\partial\boldsymbol{\kappa}}{\partial \tau_{2}}\right\}&\mathrm{Re}\left\{\frac{\partial^{H}\boldsymbol{\kappa}}{\partial \tau_{3}}\frac{\partial\boldsymbol{\kappa}}{\partial \tau_{3}}\right\}
  \end{matrix}
  \right]\nonumber\\
  &=\left[\begin{matrix}
\mathbf{F}_{\theta,\theta^{T}}&\mathbf{F}_{\theta,\boldsymbol{\alpha}^{T}}\\
\mathbf{F}^{T}_{\theta,\boldsymbol{\alpha}^{T}}&\mathbf{F}_{\boldsymbol{\alpha},\boldsymbol{\alpha}^{T}}
  \end{matrix}
  \right]=\left[\begin{matrix}
\mathbf{C}_{\theta,\theta^{T}}&\mathbf{C}_{\theta,\boldsymbol{\alpha}^{T}}\\
\mathbf{C}^{T}_{\theta,\boldsymbol{\alpha}^{T}}&\mathbf{C}_{\boldsymbol{\alpha},\boldsymbol{\alpha}^{T}}
  \end{matrix}
  \right]^{-1}=\mathbf{C}^{-1},\label{pro12}
\end{align}
\end{small}%
where $\mathbf{F}_{\theta,\theta^{T}}=\left[\begin{matrix}
\mathrm{Re}\left\{\frac{\partial^{H}\boldsymbol{\kappa}}{\partial \tau_{1}}\frac{\partial\boldsymbol{\kappa}}{\partial \tau_{1}}\right\}\end{matrix}
  \right]$,  $\mathbf{F}_{\theta,\boldsymbol{\alpha}^{T}}=\left[\begin{matrix}
\mathrm{Re}\left\{\frac{\partial^{H}\boldsymbol{\kappa}}{\partial \tau_{1}}\frac{\partial\boldsymbol{\kappa}}{\partial \tau_{2}}\right\},~\mathrm{Re}\left\{\frac{\partial^{H}\boldsymbol{\kappa}}{\partial \tau_{1}}\frac{\partial\boldsymbol{\kappa}}{\partial \tau_{3}}\right\}\end{matrix}
  \right]$ and $\mathbf{F}_{\boldsymbol{\alpha},\boldsymbol{\alpha}^{T}}=$\\
  $\left[\begin{matrix}
\mathrm{Re}\left\{\frac{\partial^{H}\boldsymbol{\kappa}}{\partial \tau_{2}}\frac{\partial\boldsymbol{\kappa}}{\partial \tau_{2}}\right\}&\mathrm{Re}\left\{\frac{\partial^{H}\boldsymbol{\kappa}}{\partial \tau_{2}}\frac{\partial\boldsymbol{\kappa}}{\partial \tau_{3}}\right\}\\
\mathrm{Re}\left\{\frac{\partial^{H}\boldsymbol{\kappa}}{\partial \tau_{3}}\frac{\partial\boldsymbol{\kappa}}{\partial \tau_{2}}\right\}&\mathrm{Re}\left\{\frac{\partial^{H}\boldsymbol{\kappa}}{\partial \tau_{3}}\frac{\partial\boldsymbol{\kappa}}{\partial \tau_{3}}\right\}
\end{matrix}
  \right]$.
According to (\ref{pro11}) and (\ref{pro12}), the CRB of $\theta$ is denoted as
\begin{small}
\begin{align}
&\mathrm{CRB}_{\theta}=\mathrm{tr}(\mathbf{C}_{\theta\theta^{T}})=\mathrm{tr}\left(\left(\mathbf{F}_{\theta\theta^{T}}-\mathbf{F}_{\theta\boldsymbol{\alpha}^{T}}\mathbf{F}_{\boldsymbol{\alpha}\boldsymbol{\alpha}^{T}}^{-1}\mathbf{F}_{\theta\boldsymbol{\alpha}^{T}}^{T}\right)^{-1}\right), \label{pro13} 
\end{align}
\end{small}%
where detailed expressions of  $\mathbf{F}_{\theta\theta^{T}}$, $\mathbf{F}_{\theta\boldsymbol{\alpha}^{T}}$ and $\mathbf{F}_{\boldsymbol{\alpha}\boldsymbol{\alpha}^{T}}$ are provided in \textbf{Appendix~\ref{appA}}.

\subsection{Problem Formulation}
In this section, we focus on the minimum transmit power problem in the RIS-assisted ISCPT NOMA system. Factors such as the sensing CRB constraint, communication rate constraints, RIS reflection unit, harvested energy constraints, and SIC decoding order constraints are taken into account to achieve the goal of reducing the transmit power. The optimization problem involves jointly optimizing the transmit beamforming matrix $\mathbf{w}_{k}$, the SIC decoding order $d(k)$, PS ratio $\rho_{k}$, and the reflection coefficient $\boldsymbol{\Theta}$ to minimize transmit power, which is formulated as
\begin{small}
\begin{subequations}
\begin{align}
\min_{\{\mathbf{w}_{k},\boldsymbol{\Theta},d(k).\rho_{k}\}}&~\sum_{k=1}^{K}\|\mathbf{w}_{k}\|^{2},\label{pro14a}\\
\mbox{s.t.}~
&\mathrm{CRB}_{\theta}\leq\epsilon,&\label{pro14b}\\
&\log_{2}(1+\mathrm{SINR}_{k})\geq \gamma_{k},&\label{pro14c}\\
&|\boldsymbol{\theta}_{n}|=1,&\label{pro14d}\\
&E_{k}\geq Q_{k},&\label{pro14e}\\
&0\leq\rho_{k}\leq1,&\label{pro14f}\\
&\mathrm{SINR}_{k}\leq \mathrm{SINR}_{\bar{k}\to k},~\text{if}~d(k)\leq d(\bar{k}),&\label{pro14g}
\end{align}\label{pro14}%
\end{subequations}
\end{small}%
where (\ref{pro14b}) is the accuracy constraint of estimating angle $\theta$. (\ref{pro14c}) is the QoS requirement. (\ref{pro14d}) is the unit-modulus constraint of the RIS reflection unit. (\ref{pro14e}) is the harvested energy constraint. (\ref{pro14f}) is the received PS ratio of the $k$-th IoT device. (\ref{pro14g}) is SIC decoding conditions.

Due to the interdependency between the optimization variables BS transmit beamforming $\{\mathbf{w}_{k}\}$, receiver PS ratio $\rho_{k}$, SIC decoding order $d(k)$, and RIS phase shift matrix $\boldsymbol{\Theta}$, and non-convex constraints (\ref{pro14b})-(\ref{pro14e}) and (\ref{pro14g}), problem (\ref{pro14}) is a non-convex problem. To address this non-convexity, we suggest employing the BCD algorithm. This algorithm involves breaking down the non-convex problem (\ref{pro14}) into several manageable sub-problems for efficient resolution. Here are the specific steps involved:
\begin{enumerate}
    \item We optimize BS transmit beamforming $\{\mathbf{w}_{k}\}$ based on the given receive PS ratio $\rho_{k}$, SIC decoding order $d(k)$, and the RIS phase shift matrix $\boldsymbol{\Theta}$.
\item We optimize the RIS phase shift matrix $\boldsymbol{\Theta}$ given the BS transmit beamforming $\{\mathbf{w}_{k}\}$, receiver PS ratio $\rho_{k}$, and the SIC decoding order $d(k)$.
\item Given the RIS phase shift matrix $\boldsymbol{\Theta}$, BS transmit beamforming $\{\mathbf{w}_{k}\}$, and receive PS ratio $\rho_{k}$, we optimize the SIC decoding order $d(k)$.
\item The receiver PS ratio $\rho_{k}$ is optimized with the given BS transmit beamforming $\{\mathbf{w}_{k}\}$, SIC decoding order $d(k)$, and RIS phase shift matrix $\boldsymbol{\Theta}$.
\end{enumerate}

\section{Proposed Algorithm for Problem (\ref{pro14})}\label{IV}
The BCD algorithm is proposed to solve the problem (\ref{pro14}) in the section. Specifically, we divide problem (\ref{pro14}) into four subproblems and propose four algorithms to solve them.
\subsection{Proposed Solution for Transmit Beamforming}
Given $\rho_{k}$, $d(k)$,~and,~$\boldsymbol{\Theta}$, problem (\ref{pro14}) is rewritten as
\begin{small}
\begin{subequations}
\begin{align}
\min_{\{\mathbf{w}_{k}\}}&~\sum_{k=1}^{K}\|\mathbf{w}_{k}\|^{2},\label{pro_15a}\\
\mbox{s.t.}~
&(\ref{pro14b}), (\ref{pro14c}),(\ref{pro14e}),~\text{and}~(\ref{pro14g}).&\label{pro_15b}
\end{align}\label{pro_15}%
\end{subequations}
\end{small}%
To deal with the non-convex constraint in (\ref{pro14b}), we introduce the auxiliary variables $\zeta$, $\zeta_{1}$, $\boldsymbol{\zeta}_{2}$, and $\boldsymbol{\zeta}_{3}$, and then let
\begin{small}
\begin{align}
&\mathbf{F}_{\theta\theta^{T}}-\mathbf{F}_{\theta\boldsymbol{\alpha}^{T}}\mathbf{F}_{\boldsymbol{\alpha}\boldsymbol{\alpha}^{T}}^{-1}\mathbf{F}_{\theta\boldsymbol{\alpha}^{T}}^{T}\geq\zeta, \zeta_{1}=\mathbf{F}_{\theta\theta^{T}}-\zeta, 
\boldsymbol{\zeta}_{2}=\mathbf{F}_{\theta\boldsymbol{\alpha}^{T}},\nonumber\\
&\boldsymbol{\zeta}_{3}=\mathbf{F}_{\boldsymbol{\alpha}\boldsymbol{\alpha}^{T}}.\label{pro15}
\end{align}
\end{small}%
Based on (\ref{pro15}), we can rewritten $\mathrm{CRB}_{\theta}\leq\epsilon$ as
\begin{small}
\begin{align}
\left[\begin{matrix}
\zeta_{1}&\boldsymbol{\zeta}_{2}\\
\boldsymbol{\zeta}_{2}^{T}&\boldsymbol{\zeta}_{3}\\
  \end{matrix}
  \right]\succeq\mathbf{0}, \frac{1}{\zeta}\leq\epsilon. \label{pro16}
\end{align}
\end{small}%
Next, we use the expressions (\ref{proA1_3})-(\ref{proA1_5}) in \textbf{Appendix~\ref{appA}}; the equality constraint conditions in (\ref{pro15}) can be equivalently transformed as in (\ref{pro17}), shown at the bottom of the next page. Then, we define $\bar{\mathbf{W}}_{k}=\mathbf{w}_{k}\mathbf{w}_{k}^{H}$. Thus, we have $\mathbf{W}\mathbf{W}^{H}=\sum_{k=1}^{K}\bar{\mathbf{W}}_{k}$,~where~$\mathrm{rank}(\bar{\mathbf{W}}_{k})=1$. Constraint condition (\ref{pro17}) is rewritten as (\ref{pro18}), and (\ref{pro18}) is given at the bottom of the next page.
\begin{figure*}[hb] 
\centering 
\hrulefill
\vspace*{6pt} 
\begin{small}
\begin{align}
&\mathrm{tr}\left(\left((\mathbf{G}^{H}\boldsymbol{\Theta}^{H}(\mathbf{h}_{s}\odot\mathbf{e})\mathbf{h}_{s}^{H}\boldsymbol{\Theta}\mathbf{G})+(\mathbf{G}^{H}\boldsymbol{\Theta}^{H}\mathbf{h}_{s}(\mathbf{h}_{s}\odot\mathbf{e})^{H}\boldsymbol{\Theta}\mathbf{G})\right)\right.\left.\mathbf{W}\mathbf{W}^{H}\left((\mathbf{G}^{H}\boldsymbol{\Theta}^{H}(\mathbf{h}_{s}\odot\mathbf{e})\mathbf{h}_{s}^{H}\boldsymbol{\Theta}\mathbf{G})+(\mathbf{G}^{H}\boldsymbol{\Theta}^{H}\mathbf{h}_{s}(\mathbf{h}_{s}\odot\mathbf{e})^{H}\right.\right.\nonumber\\
&\left.\left.\boldsymbol{\Theta}\mathbf{G})\right)^{H}\right)=\frac{\sigma_{r}^{2}(\zeta_{1}+\zeta)}{2L|\alpha|^{2}},\nonumber\\
&\mathrm{tr}\left(\mathbf{G}^{H}\boldsymbol{\Theta}^{H}\mathbf{h}_{s}\mathbf{h}_{s}^{H}\boldsymbol{\Theta}\mathbf{G}\mathbf{W}\mathbf{W}^{H}(\mathbf{G}^{H}\boldsymbol{\Theta}^{H}\mathbf{h}_{s}\mathbf{h}_{s}^{H}\boldsymbol{\Theta}\mathbf{G})^{H}\right)\mathbf{I}_{2}=\frac{\sigma_{r}^{2}\boldsymbol{\zeta}_{2}}{2L},\nonumber\\
&\mathrm{Re}\left(\mathrm{tr}\left(\mathbf{G}^{H}\boldsymbol{\Theta}^{H}\mathbf{h}_{s}\mathbf{h}_{s}^{H}\boldsymbol{\Theta}\mathbf{G}\mathbf{W}\mathbf{W}^{H}\left((\mathbf{G}^{H}\boldsymbol{\Theta}^{H}(\mathbf{h}_{s}\odot\mathbf{e})\mathbf{h}_{s}^{H}\boldsymbol{\Theta}\mathbf{G})+(\mathbf{G}^{H}\boldsymbol{\Theta}^{H}\mathbf{h}_{s}(\mathbf{h}_{s}\odot\mathbf{e})^{H}\boldsymbol{\Theta}\mathbf{G})\right)^{H}\right.\right.\left.\left.\left.(\mathbf{G}^{H}\boldsymbol{\Theta}^{H}\mathbf{h}_{s}(\mathbf{h}_{s}\odot\mathbf{e})^{H}\right.\right.\right.\nonumber\\
&\left.\left.\left.\boldsymbol{\Theta}\mathbf{G})\right)^{H}\right)[1,j]\right)=\frac{\sigma_{r}^{2}\boldsymbol{\zeta}_{3}}{2L}.\label{pro17}\\
&\mathrm{tr}\left(\left((\mathbf{G}^{H}\boldsymbol{\Theta}^{H}(\mathbf{h}_{s}\odot\mathbf{e})\mathbf{h}_{s}^{H}\boldsymbol{\Theta}\mathbf{G})+(\mathbf{G}^{H}\boldsymbol{\Theta}^{H}\mathbf{h}_{s}(\mathbf{h}_{s}\odot\mathbf{e})^{H}\boldsymbol{\Theta}\mathbf{G})\right)\right.\left.\sum_{k=1}^{K}\bar{\mathbf{W}}_{k}\left((\mathbf{G}^{H}\boldsymbol{\Theta}^{H}(\mathbf{h}_{s}\odot\mathbf{e})\mathbf{h}_{s}^{H}\boldsymbol{\Theta}\mathbf{G})+(\mathbf{G}^{H}\boldsymbol{\Theta}^{H}\mathbf{h}_{s}(\mathbf{h}_{s}\odot\mathbf{e})^{H}\right.\right.\nonumber\\
&\left.\left.\boldsymbol{\Theta}\mathbf{G})\right)^{H}\right)=\frac{\sigma_{r}^{2}(\zeta_{1}+\zeta)}{2L|\alpha|^{2}},\nonumber\\
&\mathrm{tr}\left(\mathbf{G}^{H}\boldsymbol{\Theta}^{H}\mathbf{h}_{s}\mathbf{h}_{s}^{H}\boldsymbol{\Theta}\mathbf{G}\sum_{k=1}^{K}\bar{\mathbf{W}}_{k}(\mathbf{G}^{H}\boldsymbol{\Theta}^{H}\mathbf{h}_{s}\mathbf{h}_{s}^{H}\boldsymbol{\Theta}\mathbf{G})^{H}\right)\mathbf{I}_{2}=\frac{\sigma_{r}^{2}\boldsymbol{\zeta}_{2}}{2L},\nonumber\\
&\mathrm{Re}\left(\mathrm{tr}\left(\mathbf{G}^{H}\boldsymbol{\Theta}^{H}\mathbf{h}_{s}\mathbf{h}_{s}^{H}\boldsymbol{\Theta}\mathbf{G}\sum_{k=1}^{K}\bar{\mathbf{W}}_{k}\left((\mathbf{G}^{H}\boldsymbol{\Theta}^{H}(\mathbf{h}_{s}\odot\mathbf{e})\mathbf{h}_{s}^{H}\boldsymbol{\Theta}\mathbf{G})+(\mathbf{G}^{H}\boldsymbol{\Theta}^{H}\mathbf{h}_{s}(\mathbf{h}_{s}\odot\mathbf{e})^{H}\boldsymbol{\Theta}\mathbf{G})\right)^{H}\right.\right.\left.\left.\right)[1,j]\right)=\frac{\sigma_{r}^{2}\boldsymbol{\zeta}_{3}}{2L}.\label{pro18}
\end{align}
\end{small}
\end{figure*}
It is easy to find that the constraint condition in (\ref{pro18}) is convex. Then,  the objective function in (\ref{pro_15a}) is written as
\begin{small}
\begin{align}
\sum_{k=1}^{K}\mathrm{tr}(\mathbf{W}_{k}).\label{pro19}
\end{align}
\end{small}%
Constraint condition (\ref{pro14c}) is rewritten as
\begin{small}
\begin{align}
&\frac{\rho_{k}\mathbf{h}^{H}_{u,k}\boldsymbol{\Theta}\mathbf{G}\mathbf{W}_{k}\mathbf{G}^{H}\boldsymbol{\Theta}^{H}\mathbf{h}_{u,k}}{\rho_{k}\sum_{d(j)\leq d(k)}\mathbf{h}^{H}_{u,k}\boldsymbol{\Theta}\mathbf{G}\mathbf{W}_{j}\mathbf{G}^{H}\boldsymbol{\Theta}^{H}\mathbf{h}_{u,k}+\rho_{k}\sigma_{k}^{2}+\delta_{k}^{2}}\geq 2^{\gamma_{k}}-1\nonumber\\
&\Rightarrow\rho_{k}\mathbf{h}^{H}_{u,k}\boldsymbol{\Theta}\mathbf{G}\mathbf{W}_{k}\mathbf{G}^{H}\boldsymbol{\Theta}^{H}\mathbf{h}_{u,k}\geq\rho_{k}\sum_{d(j)\leq d(k)}\mathbf{h}^{H}_{u,k}\boldsymbol{\Theta}\mathbf{G}\mathbf{W}_{j}\mathbf{G}^{H}\boldsymbol{\Theta}^{H}\nonumber\\
&\times\mathbf{h}_{u,k}+\rho_{k}\sigma_{k}^{2}+\delta_{k}^{2}(2^{\gamma_{k}}-1).\label{pro20}
\end{align}
\end{small}%
We find that the constraint condition in (\ref{pro20}) is convex over $\mathbf{W}_{k}$.
Similarly, constraint condition (\ref{pro14g}) is rewritten as
\begin{small}
\begin{align}
&\frac{\rho_{k}\mathbf{h}^{H}_{u,k}\boldsymbol{\Theta}\mathbf{G}\mathbf{W}_{k}\mathbf{G}^{H}\boldsymbol{\Theta}^{H}\mathbf{h}_{u,k}}{\rho_{k}\sum_{d(j)\leq d(k)}\mathbf{h}^{H}_{u,k}\boldsymbol{\Theta}\mathbf{G}\mathbf{W}_{j}\mathbf{G}^{H}\boldsymbol{\Theta}^{H}\mathbf{h}_{u,k}+\rho_{k}\sigma_{k}^{2}+\delta_{k}^{2}}\leq\nonumber\\
&\frac{\rho_{\bar{k}}\mathbf{h}^{H}_{u,k}\boldsymbol{\Theta}\mathbf{G}\mathbf{W}_{k}\mathbf{G}^{H}\boldsymbol{\Theta}^{H}\mathbf{h}_{u,\bar{k}}}{\rho_{k}\sum_{d(j)\leq d(k)}\mathbf{h}^{H}_{u,\bar{k}}\boldsymbol{\Theta}\mathbf{G}\mathbf{W}_{j}\mathbf{G}^{H}\boldsymbol{\Theta}^{H}\mathbf{h}_{u,\bar{k}}+\rho_{k}\sigma_{k}^{2}+\delta_{k}^{2}}. \label{pro21}
\end{align}
\end{small}%
Since the constraint condition in (\ref{pro21}) is still non-convex, we use the SCA method to approximate (\ref{pro21}), and we have
\begin{small}
\begin{align}
&\bar{f}_{1}(\mathbf{W}_{k})-\ln\left(\sum_{d(j)>d(k)}\mathbf{h}^{H}_{u,k}\boldsymbol{\Theta}\mathbf{G}\mathbf{W}_{j}\mathbf{G}^{H}\boldsymbol{\Theta}^{H}\mathbf{h}_{u,k}+A_{k}\right)\nonumber\\
&-\ln(\mathbf{h}^{H}_{u,k}\boldsymbol{\Theta}\mathbf{G}\mathbf{W}_{k}\mathbf{G}^{H}\boldsymbol{\Theta}^{H}\mathbf{h}_{u,\bar{k}})+\bar{f}_{2}(\mathbf{W}_{j})\leq 0, \label{pro22}
\end{align}
\end{small}%
where 
\begin{small}
\begin{align}
&\bar{f}_{1}(\mathbf{W}_{k})=\ln(\rho_{k}\mathbf{h}^{H}_{u,k}\boldsymbol{\Theta}\mathbf{G}\mathbf{W}_{k}^{0}\mathbf{G}^{H}\boldsymbol{\Theta}^{H}\mathbf{h}_{u,k})\nonumber\\
&+\mathrm{tr}\left((\frac{\rho_{k}\mathbf{h}^{H}_{u,k}\mathbf{h}_{u,k}}{\rho_{k}\mathbf{h}^{H}_{u,k}\boldsymbol{\Theta}\mathbf{G}\mathbf{W}_{k}^{0}\mathbf{G}^{H}\boldsymbol{\Theta}^{H}\mathbf{h}_{u,k}})^{H}(\mathbf{W}_{k}-\mathbf{W}_{k}^{0})\right),\nonumber\\
&\bar{f}_{2}(\mathbf{W}_{j})=\ln(\rho_{k}\mathbf{h}^{H}_{u,\bar{k}}\boldsymbol{\Theta}\mathbf{G}\mathbf{W}_{j}^{0}\mathbf{G}^{H}\boldsymbol{\Theta}^{H}\mathbf{h}_{u,\bar{k}})\nonumber\\
&+\mathrm{tr}\left((\frac{\rho_{k}\mathbf{h}^{H}_{u,\bar{k}}\mathbf{h}_{u,\bar{k}}}{\rho_{k}\mathbf{h}^{H}_{u,\bar{k}}\boldsymbol{\Theta}\mathbf{G}\mathbf{W}_{j}^{0}\mathbf{G}^{H}\boldsymbol{\Theta}^{H}\mathbf{h}_{u,\bar{k}}})^{H}(\mathbf{W}_{j}-\mathbf{W}_{j}^{0})\right).\label{pro_22}
\end{align}
\end{small}%
The constraint condition in (\ref{pro22}) is convex. Next, we continue with the harvested energy constraint in (\ref{pro14e}), and it is rewritten as
\begin{small}
\begin{align}
\eta_{k}(1-\rho_{k})(\mathbf{h}^{H}_{u,k}\boldsymbol{\Theta}\mathbf{G}\mathbf{W}_{k}\mathbf{G}^{H}\boldsymbol{\Theta}^{H}\mathbf{h}_{u,k}+\sigma_{k}^{2})\geq Q_{k}. \label{pro23}
\end{align}
\end{small}%
The constraint condition in (\ref{pro23}) is convex.
Based on the above consideration, problem (\ref{pro_15}) is rewritten as
\begin{small}
\begin{subequations}
\begin{align}
\min_{\{\mathbf{W}_{k}\}}&~\sum_{k=1}^{K}\mathrm{tr}(\mathbf{W}_{k}),\label{pro24a}\\
\mbox{s.t.}~
&(\ref{pro16}), (\ref{pro18}),(\ref{pro20}), (\ref{pro22}), (\ref{pro23})&\label{pro24b}\\
&\mathrm{rank}(\mathbf{W}_{k})=1,&\label{pro24c}
\end{align}\label{pro24}%
\end{subequations}
\end{small}%
The problem in (\ref{pro24}) is an Semidefinite programming (SDP) problem, and we use SDR to solve it. Also, we can prove that the optimal solution of the problem (\ref{pro24}) also satisfies $\mathrm{rank}(\mathbf{W}_{k})=1$. The proof is given in \textbf{Appendix~\ref{appB}}.

\subsection{Proposed Algorithm for RIS Phase Shift Optimization Problem}
Given $\mathbf{w}_{k}$, $d(k)$,~and~$\rho_{k}$, the RIS phase shift optimization problem is given by
\begin{small}
\begin{subequations}
\begin{align}
\min_{\{\boldsymbol{\Theta}\}}&~\sum_{k=1}^{K}\|\mathbf{w}_{k}\|^{2},\label{pro25a}\\
\mbox{s.t.}~
&(\ref{pro14b}),(\ref{pro14c}),(\ref{pro14d}), (\ref{pro14e}),(\ref{pro14g}).&\label{pro25b}
\end{align}\label{pro25}%
\end{subequations}
\end{small}%
The constraint condition in (\ref{pro14b}) is reformulated as
\begin{small}
\begin{align}
&\left[\begin{matrix}
\frac{2L|\alpha_{t}|^{2}}{\sigma_{r}^{2}}\mathrm{Re}(\bar{g}_{1}(\boldsymbol{\theta}))-J&\frac{2L}{\sigma_{r}^{2}}\mathrm{Re}\left(\alpha_{t}^{*}\left[\begin{matrix}
\bar{g}_{2}(\boldsymbol{\theta})\\
\bar{g}_{3}(\boldsymbol{\theta})\\
  \end{matrix}\right][1~j]\right)\\
\frac{2L}{\sigma_{r}^{2}}\mathrm{Re}\left(\alpha_{t}^{*}\left[\begin{matrix}
1\\
j\\
  \end{matrix}\right]\left[\begin{matrix}
\bar{g}_{2}(\boldsymbol{\theta})&
\bar{g}_{3}(\boldsymbol{\theta})\\
  \end{matrix}\right]\right)&\frac{2L}{\sigma_{r}^{2}}\bar{g}_{4}(\boldsymbol{\theta})\mathbf{I}_{2}\\
  \end{matrix}
  \right]\nonumber\\
  &\succeq\mathbf{0}, \label{pro26}
\end{align}
\end{small}%
where $\bar{g}_{1}(\boldsymbol{\theta})=g_{1}(\boldsymbol{\theta})+g_{2}(\boldsymbol{\theta})+g_{3}(\boldsymbol{\theta})+g_{4}(\boldsymbol{\theta})=(\boldsymbol{\theta}\otimes\boldsymbol{\theta})^{H}\hat{\boldsymbol{\Omega}}_{1}(\boldsymbol{\theta}\otimes\boldsymbol{\theta})$,
$\bar{g}_{2}(\boldsymbol{\theta})=(\boldsymbol{\theta}\otimes\boldsymbol{\theta})^{H}\hat{\boldsymbol{\Omega}}_{2}(\boldsymbol{\theta}\otimes\boldsymbol{\theta})$,
$\bar{g}_{3}(\boldsymbol{\theta})=(\boldsymbol{\theta}\otimes\boldsymbol{\theta})^{H}\hat{\boldsymbol{\Omega}}_{3}(\boldsymbol{\theta}\otimes\boldsymbol{\theta})$,
$\bar{g}_{4}(\boldsymbol{\theta})=(\boldsymbol{\theta}\otimes\boldsymbol{\theta})^{H}\hat{\boldsymbol{\Omega}}_{4}(\boldsymbol{\theta}\otimes\boldsymbol{\theta})$.
By introducing auxiliary variables $\{\xi_{1},\xi_{2},\xi_{3},\xi_{4},\xi_{5},\xi_{6}\}$, the SINR constraint and the SIC constraint can be reformulated as
\begin{small}
\begin{align}
&g_{7}(\boldsymbol{\theta})-g_{8}(\boldsymbol{\theta})=\xi_{5}-\xi_{6}\leq\boldsymbol{\theta}^{H}\tilde{\mathbf{D}}_{k,k}\boldsymbol{\theta},\label{pro27}\\
&\bar{g}_{1}(\boldsymbol{\theta})=\xi_{1},\bar{g}_{2}(\boldsymbol{\theta})=\xi_{2},\bar{g}_{3}(\boldsymbol{\theta})=\xi_{3},\bar{g}_{4}(\boldsymbol{\theta})=\xi_{4},g_{7}(\boldsymbol{\theta})=\xi_{5},\nonumber\\
&g_{8}(\boldsymbol{\theta})=\xi_{6}.\label{pro28}
\end{align}
\end{small}%
The proof is given in \textbf{Appendix~\ref{appC}}. Next,
we use the SCA algorithm to get a lower bound of $\boldsymbol{\theta}^{H}\tilde{\mathbf{D}}_{k,k}\boldsymbol{\theta}$ and it is given
\begin{small}
\begin{align}
\xi_{5}-\xi_{6} \leq 2\mathrm{Re}\{\boldsymbol{\theta}^{t,H}\tilde{\mathbf{D}}_{k,k}\boldsymbol{\theta}\}-\boldsymbol{\theta}^{t,H}\tilde{\mathbf{D}}_{k,k}\boldsymbol{\theta}^{t}\leq \boldsymbol{\theta}^{H}\tilde{\mathbf{D}}_{k,k}\boldsymbol{\theta}. \label{pro29}
\end{align}
\end{small}%
Similarly, constraint (\ref{pro14e}) is approximated as
\begin{small}
\begin{align}
\frac{Q_{k}}{\eta_{k}(1-\rho_{k})}-\sigma_{k}^{2}\leq 2\mathrm{Re}\{\boldsymbol{\theta}^{t,H}\tilde{\mathbf{D}}_{k,j}\boldsymbol{\theta}\}-\boldsymbol{\theta}^{t,H}\tilde{\mathbf{D}}_{k,k}\boldsymbol{\theta}^{t}. \label{pro30}
\end{align}
\end{small}%
Problem (\ref{pro25}) is rewritten as
\begin{small}
\begin{subequations}
\begin{align}
\min_{\boldsymbol{\Theta},\xi_{1},\xi_{2},\atop{\xi_{3},\xi_{4},\atop{\xi_{5},\xi_{6}}}}&~\sum_{k=1}^{K}\|\mathbf{w}_{k}\|^{2},\label{pro31a}\\
\mbox{s.t.}~
&
\left[\begin{matrix}
\frac{2L|\alpha_{t}|^{2}}{\sigma_{r}^{2}}\mathrm{Re}(\xi_{1})-J&\frac{2L}{\sigma_{r}^{2}}\mathrm{Re}\left(\alpha_{t}^{*}\left[\begin{matrix}
\xi_{2}\\
\xi_{3}\\
  \end{matrix}\right][1~j]\right)\\
\frac{2L}{\sigma_{r}^{2}}\mathrm{Re}\left(\alpha_{t}^{*}\left[\begin{matrix}
1\\
j\\
  \end{matrix}\right]\left[\begin{matrix}
\xi_{2}&
\xi_{3}\\
  \end{matrix}\right]\right)&\frac{2L}{\sigma_{r}^{2}}\xi_{4}\mathbf{I}_{2}\\
  \end{matrix}
  \right]\succeq\mathbf{0},&\label{pro31_b}\\
&(\ref{pro28}),(\ref{pro29}),(\ref{pro30}),(\ref{pro14e}).&\label{pro31b}
\end{align}\label{pro31}%
\end{subequations}
\end{small}

According to\cite{b35}, the augmented Lagrangian function of problem (\ref{pro31}) is denoted as
\begin{small}
\begin{subequations}
\begin{align}
\min_{\{\boldsymbol{\Theta},\{\xi_{1},\xi_{2},\xi_{3},\xi_{4},\xi_{5},\xi_{6}\}\}}&~\frac{1}{2\rho_{1}}\sum_{k=1}^{6}|\bar{g}_{k}(\boldsymbol{\theta})-\xi_{k}+\bar{\rho}_{1}\bar{\xi}_{k}|^{2}+\sum_{k=1}^{K}\|\mathbf{w}_{k}\|^{2},\label{pro32a}\\
\mbox{s.t.}~
&(\ref{pro31_b}),(\ref{pro29}),(\ref{pro30}),(\ref{pro14e}),&\label{pro32b}
\end{align}\label{pro32}%
\end{subequations}
\end{small}%
where $\{\bar{\xi}_{1},\bar{\xi}_{2},\bar{\xi}_{3},\bar{\xi}_{4},\bar{\xi}_{5},\bar{\xi}_{6}\}$ are the dual variables and $\rho_{1}>0$ is the penalty factor. To solve the problem in (\ref{pro32}), we use the alternating optimization (AO) method. Based on the AO method, problem (\ref{pro32}) can be divided into two non-convex subproblems 
\begin{small}
\begin{subequations}
\begin{align}
\min_{\{\xi_{1},\xi_{2},\xi_{3},\xi_{4},\xi_{5},\xi_{6}\}}&~\sum_{k=1}^{K}\|\mathbf{w}_{k}\|^{2}+\frac{1}{2\rho_{1}}\sum_{k=1}^{6}|\bar{g}_{k}(\boldsymbol{\theta})-\xi_{k}+\bar{\rho}_{1}\bar{\xi}_{k}|^{2},\label{pro_32a}\\
\mbox{s.t.}~
&(\ref{pro31_b}),&\label{pro_32b}
\end{align}\label{pro_32}%
\end{subequations}
\end{small}%
and
\begin{small}
\begin{subequations}
\begin{align}
\min_{\{\boldsymbol{\Theta}\}}&~\sum_{k=1}^{K}\|\mathbf{w}_{k}\|^{2}+\frac{1}{2\rho_{1}}\sum_{k=1}^{6}|\bar{g}_{k}(\boldsymbol{\theta})-\xi_{k}+\bar{\rho}_{1}\bar{\xi}_{k}|^{2},\label{pro__32a}\\
\mbox{s.t.}~
&(\ref{pro29}),(\ref{pro30}),(\ref{pro14e}).&\label{pro__32b}
\end{align}\label{pro__32}%
\end{subequations}
\end{small}%
Subsequently, we transform the problem in (\ref{pro_32}) as 
\begin{small}
\begin{subequations}
\begin{align}
\min_{\{\xi_{1},\xi_{2},\xi_{3},\xi_{4},\xi_{5},\xi_{6}\}}&~\frac{1}{2\rho_{1}}\sum_{k=1}^{6}\xi_{k}\xi_{k}^{H}+(\bar{g}_{k}(\boldsymbol{\theta})-\bar{\rho}_{1}\bar{\xi}_{k})(\bar{g}_{k}(\boldsymbol{\theta})-\bar{\rho}_{1}\bar{\xi}_{k})^{H}\nonumber\\
&-2\mathrm{Re}{\xi_{k}(\bar{g}_{k}(\boldsymbol{\theta})-\bar{\rho}_{1}\bar{\xi}_{k})^{H}},\label{pro__32_a}\\
\mbox{s.t.}~
&(\ref{pro31_b}), &\label{pro__32_b}
\end{align}\label{pro__32_}%
\end{subequations}
\end{small}%
where $(\bar{g}_{k}(\boldsymbol{\theta})-\bar{\rho}_{1}\bar{\xi}_{k})(\bar{g}_{k}(\boldsymbol{\theta})-\bar{\rho}_{1}\bar{\xi}_{k})^{H}$ and optimization variables $\{\xi_{1},\xi_{2},\xi_{3},\xi_{4},\xi_{5},\xi_{6}\}$ are unrelated. Thus, problem (\ref{pro__32_}) is re-expressed as
\begin{small}
\begin{subequations}
\begin{align}
\min_{\{\xi_{1},\xi_{2},\xi_{3},\xi_{4},\xi_{5},\xi_{6}\}}&~\frac{1}{2\rho_{1}}\sum_{k=1}^{6}\xi_{k}\xi_{k}^{H}-2\mathrm{Re}{\xi_{k}(\bar{g}_{k}(\boldsymbol{\theta})-\bar{\rho}_{1}\bar{\xi}_{k})^{H}},\label{pro__32_a1}\\
\mbox{s.t.}~
&(\ref{pro31_b}). &\label{pro__32_b1}
\end{align}\label{pro__32_1}%
\end{subequations}
\end{small}%
According to \cite{b36}, problem (\ref{pro__32_1}) is a standard SDP problem, and it can be solved by using the CVX tool.
Then,  given $\{\xi_{1},\xi_{2},\xi_{3},\xi_{4},\xi_{5},\xi_{6}\}$, we solve problem (\ref{pro__32_1}). Next, to deal with the non-convex constraints in (\ref{pro14e}), we introduce the auxiliary variables $\boldsymbol{\vartheta}_{1},\boldsymbol{\vartheta}_{2},~\text{and}~\boldsymbol{\vartheta}_{3}$, and then (\ref{pro__32}) is rewritten as
\begin{small}
\begin{subequations}
\begin{align}
\min_{\boldsymbol{\Theta}}&~\sum_{k=1}^{K}\|\mathbf{w}_{k}\|^{2}+\frac{1}{2\rho_{1}}\sum_{k=1}^{6}|\bar{g}_{k}(\boldsymbol{\theta},\boldsymbol{\vartheta}_{1},\boldsymbol{\vartheta}_{2},\boldsymbol{\vartheta}_{3})-\xi_{k}+\bar{\rho}_{1}\bar{\xi}_{k}|^{2},\label{pro34a}\\
\mbox{s.t.}~
&(\ref{pro29}),(\ref{pro30}),(\ref{pro14e}),&\label{pro34b}\\
&\boldsymbol{\vartheta}_{1}=\boldsymbol{\theta}, \boldsymbol{\vartheta}_{2}=\boldsymbol{\theta}\otimes\boldsymbol{\vartheta}_{1}, \boldsymbol{\vartheta}_{3}=\boldsymbol{\vartheta}_{2},&\label{pro34c}\\
&|\boldsymbol{\vartheta}_{1,n}|=|\boldsymbol{\theta}_{n}|=1, |\boldsymbol{\vartheta}_{2,j}|=|\boldsymbol{\vartheta}_{3,j}|=1.&\label{pro34d}
\end{align}\label{pro34}%
\end{subequations}
\end{small}%
Problem (\ref{pro34}) is equivalently rewritten as
\begin{small}
\begin{subequations}
\begin{align}
\min_{\boldsymbol{\Theta}}&~\sum_{k=1}^{K}\|\mathbf{w}_{k}\|^{2}+\frac{1}{2\lambda_{1}}\sum_{k=1}^{6}|\bar{g}_{k}(\boldsymbol{\theta},\boldsymbol{\vartheta}_{1},\boldsymbol{\vartheta}_{2},\boldsymbol{\vartheta}_{3})+\hat{\xi}_{k}|^{2}\nonumber\\
&+\frac{1}{2\lambda_{2}}\|\boldsymbol{\vartheta}_{1}-\boldsymbol{\theta}+\lambda_{2}\boldsymbol{\mu}_{1}\|^{2}+\frac{1}{2\lambda_{3}}\|\boldsymbol{\vartheta}_{2}-\boldsymbol{\theta}\otimes\boldsymbol{\vartheta}_{1}+\lambda_{3}\boldsymbol{\mu}_{2}\|^{2}\nonumber\\
&+\frac{1}{2\lambda_{4}}\|\boldsymbol{\vartheta}_{3}-\boldsymbol{\vartheta}_{2}+\lambda_{4}\boldsymbol{\mu}_{3}\|^{2},\label{pro35a}\\
\mbox{s.t.}~
&(\ref{pro29}),(\ref{pro30}),(\ref{pro14e}), (\ref{pro34d}),&\label{pro35b}
\end{align}\label{pro36}%
\end{subequations}
\end{small}%
where $\hat{\xi}_{k}=-\xi_{k}+\bar{\rho}_{1}\bar{\xi}_{k}$.
We transform $\bar{g}_{k}(\boldsymbol{\theta},\boldsymbol{\vartheta}_{1},\boldsymbol{\vartheta}_{2},\boldsymbol{\vartheta}_{3})+\hat{\xi}_{k}$ into a linear function and it can be rewritten as
\begin{small}
\begin{align}
&\bar{g}_{k}(\boldsymbol{\theta},\boldsymbol{\vartheta}_{1},\boldsymbol{\vartheta}_{2},\boldsymbol{\vartheta}_{3})+\hat{\xi}_{k}=\hat{s}_{\boldsymbol{\theta},k}+\boldsymbol{\theta}^{H}\mathbf{d}_{\boldsymbol{\theta},k}=\hat{s}_{\boldsymbol{\vartheta}_{1},k}+\mathbf{d}_{\boldsymbol{\vartheta}_{1},k}^{H}\boldsymbol{\vartheta}_{1}\nonumber\\
&=\hat{s}_{\boldsymbol{\vartheta}_{2},k}+\mathbf{d}_{\boldsymbol{\vartheta}_{2},k}^{H}\boldsymbol{\vartheta}_{2}=\hat{s}_{\boldsymbol{\vartheta}_{3},k}+\mathbf{d}_{\boldsymbol{\vartheta}_{1},3}^{H}\boldsymbol{\vartheta}_{3}.\label{pro_36}
\end{align}
\end{small}%
In addition, $\boldsymbol{\theta}\otimes\boldsymbol{\psi}_{1}$ is rewritten as
\begin{small}
\begin{align}
\boldsymbol{\theta}\otimes\boldsymbol{\psi}_{1}=\mathrm{vec}\{\boldsymbol{\psi}_{1}\boldsymbol{\theta}^{T}\}=(\boldsymbol{\theta}\otimes\mathbf{I}_{N})\boldsymbol{\psi}_{1}=(\mathbf{I}_{N}\otimes\boldsymbol{\psi}_{1})\boldsymbol{\theta}.\label{pro37}
\end{align}
\end{small}%
According to (\ref{pro37}), (\ref{pro35a}) is rewritten as
\begin{small}
\begin{subequations}
\begin{align}
\min_{\boldsymbol{\theta}}&~\boldsymbol{\theta}^{H}\boldsymbol{\Psi}_{\boldsymbol{\theta}}\boldsymbol{\theta}+\mathbf{Re}(\boldsymbol{\psi}_{\boldsymbol{\theta}}^{H}\boldsymbol{\theta}),\label{pro38a}\\
\mbox{s.t.}~
&(\ref{pro29}),(\ref{pro30}),(\ref{pro14e}), (\ref{pro34d}),&\label{pro38b}
\end{align}\label{pro38}%
\end{subequations}
\end{small}%
where
$\boldsymbol{\Psi}_{\boldsymbol{\theta}}=\frac{1}{2\lambda_{1}}\sum_{k=1}^{6}\mathbf{d}_{\boldsymbol{\theta},k}\mathbf{d}_{\boldsymbol{\theta},k}^{H}
$, $\boldsymbol{\psi}_{\boldsymbol{\theta}}=\frac{1}{\lambda_{1}}\sum_{k=1}^{6}\hat{s}_{\boldsymbol{\theta},k}^{*}\mathbf{d}_{\boldsymbol{\theta},k}+\boldsymbol{\mu}_{1}-\frac{\boldsymbol{\vartheta}_{1}}{\lambda_{2}}-(\mathbf{I}_{N}\otimes\boldsymbol{\vartheta}_{1}^{H})(\boldsymbol{\mu}_{2}+\frac{\boldsymbol{\vartheta}_{2}}{\lambda_{3}})$.
Then, we use MM algorithm to solve problem (\ref{pro38}) and the upperbound of $\boldsymbol{\theta}^{H}\boldsymbol{\Psi}_{\boldsymbol{\theta}}\boldsymbol{\theta}$ is expressed as
\begin{small}
\begin{align}
&\boldsymbol{\theta}^{H}\boldsymbol{\Psi}_{\boldsymbol{\theta}}\boldsymbol{\theta}\leq N\|\boldsymbol{\Psi}_{\boldsymbol{\theta}}\|_{F}^{2}+2\mathrm{Re}(\boldsymbol{\theta}^{t,H}(\boldsymbol{\Psi}_{\boldsymbol{\theta}}-\|\boldsymbol{\Psi}_{\boldsymbol{\theta}}\|_{F}^{2}\mathbf{I}_{N})\boldsymbol{\theta})\nonumber\\
&+\boldsymbol{\theta}^{t,H}(\|\boldsymbol{\Psi}_{\boldsymbol{\theta}}\|_{F}^{2}\mathbf{I}_{N}-\boldsymbol{\Psi}_{\boldsymbol{\theta}})\boldsymbol{\theta}^{t}.\label{pro39}
\end{align}
\end{small}%
Similarly, we use the process of problem (\ref{pro38}) to update $\boldsymbol{\vartheta}_{1}$, $\boldsymbol{\vartheta}_{2}$ and $\boldsymbol{\vartheta}_{3}$.
Finally, we update $\bar{\xi}_{k}$, $\boldsymbol{\mu}_{1}$, $\boldsymbol{\mu}_{2}$, $\boldsymbol{\mu}_{3}$ and the expression is given by
\begin{small}
\begin{align}
&\bar{\xi}_{k}=\bar{\xi}_{k}+(\hat{\xi}_{k}-\xi_{k})/\bar{\rho}_{k}, \boldsymbol{\mu}_{1}=\boldsymbol{\mu}_{1}+\frac{(\boldsymbol{\theta}-\boldsymbol{\vartheta}_{1})}{\lambda_{2}},\nonumber\\
&\boldsymbol{\mu}_{2}=\boldsymbol{\mu}_{2}+\frac{(\boldsymbol{\vartheta}_{2}-\boldsymbol{\theta}\otimes\boldsymbol{\vartheta}_{1})}{\lambda_{3}},~\text{and}~\boldsymbol{\mu}_{3}=\boldsymbol{\mu}_{3}+\frac{(\boldsymbol{\vartheta}_{3}-\boldsymbol{\vartheta}_{2})}{\lambda_{4}}.\label{po40}
\end{align}
\end{small}%

\subsection{Proposed Algorithm for SIC Decoding Order Optimization Problem}
In this section, we propose an algorithm to determine the SIC decoding order based on integrated channel gains. NOMA is an important research of this paper, making a reasonable SIC decoding order crucial\cite{b13}. The SIC decoding order is determined by the channel gain from the BS to each IoT device, considering the impact of the RIS\cite{b15}. The introduction of the RIS affects not only the direct channel gain from the BS to the ground IoT device but also the integrated channel gain from the BS to all IoT devices. Changes in the RIS phase shift matrix influence these integrated channel gains\cite{b16}. Since the same phase shift affects different IoT devices inconsistently, the integrated channel gains of all devices cannot be maximized simultaneously. Therefore, we aim to maximize the sum of the integrated channel gains from the BS to all IoT devices by optimizing the RIS phase shift. Sorting the integrated channel gains of all IoT devices helps determine the SIC decoding order. The maximization of the integrated channel gains of all IoT devices can be formulated as 
\begin{small}
\begin{subequations}
\begin{align}
\max_{\boldsymbol{\Theta}}&~\sum_{k=1}^{K}\|\mathbf{h}_{u,k}^{H}\boldsymbol{\Theta}\mathbf{G}\|^{2},\label{pro41a}\\
\mbox{s.t.}~
&(\ref{pro14d}),&\label{pro41b}
\end{align}\label{pro41}%
\end{subequations}
\end{small}%
problem (\ref{pro41}) is rewritten as
\begin{small}
\begin{subequations}
\begin{align}
\max_{\boldsymbol{\theta}}&~\sum_{k=1}^{K}\boldsymbol{\theta}^{H}\underbrace{\mathrm{diag}(\mathbf{h}_{u,k}^{H})\mathbf{G}\mathbf{G}^{H}\mathrm{diag}(\mathbf{h}_{u,k})}_{\boldsymbol{\Lambda}_{k}}\boldsymbol{\theta},\label{pro42a}\\
\mbox{s.t.}~
&(\ref{pro14d}),&\label{pro42b}
\end{align}\label{pro42}%
\end{subequations}
\end{small}%
and we use the MM algorithm to get an upper bound of
\begin{small}
\begin{align}
&\sum_{k=1}^{K}\boldsymbol{\theta}^{H}\boldsymbol{\Lambda}_{k}\boldsymbol{\theta}\leq N\|\boldsymbol{\Lambda}_{k}\|_{F}^{2}+2\mathrm{Re}(\boldsymbol{\theta}^{t,H}(\boldsymbol{\Lambda}_{k}-\|\boldsymbol{\Lambda}_{k}\|_{F}^{2}\mathbf{I}_{N})\boldsymbol{\theta})\nonumber\\
&+\boldsymbol{\theta}^{t,H}(\|\boldsymbol{\Lambda}_{k}\|_{F}^{2}\mathbf{I}_{N}-\boldsymbol{\Lambda}_{k})\boldsymbol{\theta}^{t}.\label{pro_43}
\end{align}
\end{small}%
Problem (\ref{pro41}) is approximated as
\begin{small}
\begin{subequations}
\begin{align}
\max_{\boldsymbol{\theta}}&~\sum_{k=1}^{K}2\mathrm{Re}(\boldsymbol{\theta}^{t,H}(\boldsymbol{\Lambda}_{k}-\|\boldsymbol{\Lambda}_{k}\|_{F}^{2}\mathbf{I}_{N})\boldsymbol{\theta}),\label{pro43a}\\
\mbox{s.t.}~
&(\ref{pro14d}).&\label{pro43b}
\end{align}\label{pro43}%
\end{subequations}
\end{small}%
Hence, the optimal solution of problem (\ref{pro43}) is given as
\begin{align}
&\boldsymbol{\theta}^{*}=e^{j(\angle \boldsymbol{\Lambda})}, \boldsymbol{\Lambda}=\sum_{k=1}^{K}2\boldsymbol{\theta}^{t,H}(\boldsymbol{\Lambda}_{k}-\|\boldsymbol{\Lambda}_{k}\|_{F}^{2}\mathbf{I}_{N}), \label{pro44}
\end{align}
and we can obtain the one that maximizes
the combined channel gain of all IoT devices as 
\begin{align}
t^{*}=\arg\max_{t}\sum_{k=1}^{K}\|\mathbf{h}_{u,k}^{H}\mathrm{diag}(\boldsymbol{\theta}^{*})\mathbf{G}\|^{2}.\label{pro45}
\end{align}

\subsection{Proposed Algorithm for PS Ratio Optimization Problem}
According to the decoding order of SIC is determined in the initial phase, considering the provided BS beamforming vector and RIS phase shift, problem (\ref{pro14}) can be converted into a feasibility-check problem (\ref{pro46}), which can be expressed as
\begin{subequations}
\begin{align}
\mathrm{find}_{\{\rho_{k}\}}&\label{pro46a}\\
\mbox{s.t.}~
&(\ref{pro14c}),(\ref{pro14e}),(\ref{pro14f}),(\ref{pro14g}).&\label{pro46b}
\end{align}\label{pro46}%
\end{subequations}
Converting the non-convex constraint (\ref{pro14f}) in problem (\ref{pro46}) using the SCA method results in a non-convex optimization problem
\begin{align}
&\frac{|\mathbf{h}^{H}_{u,k}\boldsymbol{\Theta}\mathbf{G}\mathbf{w}_{k}|^{2}\delta_{\bar{k}}^{2}}{\rho_{\bar{k}}}-|\mathbf{h}^{H}_{u,\bar{k}}\boldsymbol{\Theta}\mathbf{G}\mathbf{w}_{k}|^{2}\delta_{\bar{k}}^{2}\left(\frac{1}{\rho_{k}^{t}}-\frac{1}{(\rho_{k}^{t})^{2}}(\rho_{k}-\rho_{k}^{t})\right)\nonumber\\
&\leq |\mathbf{h}^{H}_{u,\bar{k}}\boldsymbol{\Theta}\mathbf{G}\mathbf{w}_{k}|^{2}\sum_{d(j)\leq d(k)}\left(|\mathbf{h}^{H}_{u,k}\boldsymbol{\Theta}\mathbf{G}\mathbf{w}_{j}|^{2}+\sigma_{k}^{2}\right)\nonumber\\
&-|\mathbf{h}^{H}_{u,k}\boldsymbol{\Theta}\mathbf{G}\mathbf{w}_{k}|^{2}|\mathbf{h}^{H}_{u,j}\boldsymbol{\Theta}\mathbf{G}\mathbf{w}_{\bar{k}}|^{2}, if~d(k)<d(\bar{k}).\label{pro47}
\end{align}
The problem (\ref{pro47}) is convex, and it is able to be resolved using CVX \cite{b35}.
\begin{algorithm}%
\caption{Proposed BCD Algorithm for Problem (\ref{pro14})} \label{algo1}
\hspace*{0.02in}{\bf Initialize:}
$\mathbf{w}_{k}^{(0)}$, $\boldsymbol{\Theta}^{(0)}$, $d(k)^{(0)}$, $\rho_{k}^{(0)}$.\\
\hspace*{0.02in}{\bf Repeat:}~$t=t+1$.\\
Given $\boldsymbol{\Theta}^{(0)}$, $d(k)^{(0)}$, $\rho_{k}^{(0)}$, utilizing SCA and SDR for problem (\ref{pro_15}) and obtaining solution $\mathbf{w}_{k}^{*}$ of problem (\ref{pro_15}). Update $\mathbf{w}_{k}^{(t+1)}=\mathbf{w}_{k}^{*}$;\\
Given $\mathbf{w}_{k}^{(t+1)}$, $d(k)^{(0)}$, $\rho_{k}^{(0)}$, utilizing ADMM for problem (\ref{pro25}) and obtaining solution $\boldsymbol{\Theta}^{*}$ of problem (\ref{pro25}). Update $\boldsymbol{\Theta}^{(t+1)}=\boldsymbol{\Theta}^{*}$;\\
Given $\mathbf{w}_{k}^{(t+1)}$, $\boldsymbol{\Theta}^{(t+1)}$, $\rho_{k}^{(0)}$, utilizing (\ref{pro44}) for problem (\ref{pro41}) and obtaining solution of problem (\ref{pro41}). Update $d(k)^{(t+1)}=d(k)^{*}$;\\
Given $\mathbf{w}_{k}^{(t+1)}$, $\boldsymbol{\Theta}^{(t+1)}$, $d(k)^{(t+1)}$, utilizing SCA for problem (\ref{pro46}) and obtaining solution $\rho_{k}^{(*)}$ of problem (\ref{pro46}). Update $\rho_{k}^{(t+1)}=\rho_{k}^{(*)}$;\\
\hspace*{0.02in}{\bf Until:}~$|\sum_{k=1}^{K}\|\mathbf{w}_{k}^{(t+1)}\|^{2}-\sum_{k=1}^{K}\|\mathbf{w}_{k}^{(t)}\|^{2}|\leq \nu$.\\
\hspace*{0.02in}{\bf Output:}
$\mathbf{w}_{k}^{(t+1)}$, $\boldsymbol{\Theta}^{(t+1)}$, $d(k)^{(t+1)}$, $\rho_{k}^{(t+1)}$.\\
\end{algorithm}

\subsection{Convergence Analysis}
In this section, to analyze the convergence of \textbf{Algorithm~\ref{algo1}}. The number of iterations of \textbf{Algorithm~\ref{algo1}} is set as $t$. The computational complexity of problem (\ref{pro_15}) is $\mathcal{O}((N_{T}+1)^{3.5})$. Based on the ADMM method in\cite{b37}, computational complexity of problem (\ref{pro25}) is $\mathcal{O}(k_{ADMM}M^{3})$, where $k_{ADMM}$ is the number of iterations. The computational complexity of problem (\ref{pro41}) is $\mathcal{O}(M)$. The computational complexity of problem (\ref{pro46}) is $\mathcal{O}((K)^{3.5})$. In summary, the computational complexity of \textbf{Algorithm~\ref{algo1}} is $\mathcal{O}(t((N_{T}+1)^{3.5}+k_{ADMM}M^{3}+M+(K)^{3.5}))$.

\section{Numerical Results}\label{V}
In this part, the performance of the proposed algorithm in RIS-aided ISCPT system based on NOMA is analyzed using simulation results.  In the coordinate system, the position of BS is $(0m, 0m,
15m)$, where $15m$ represents the height of the BS. The RIS is positioned on a building and it is set as $(25m, 25m, 15m)$. There are $4$ IoT devices, which are randomly distributed on a disk of radius radius of $50m$. Additionally, there is a $1$ target and it is randomly distributed in a circle with a radius $30m$ centered at $(0m,40m,0)$. The paper assumes that the RIS comprises $64$ reflecting elements. Furthermore, we assume that the path loss exponents are set as follows: $\alpha_{G}= 3$, $\alpha_{u} = 2.2$, and the target reflection coefficient is $\alpha=0.01$. The noise parameters $\sigma^{2}=-70$~dBm, and $\eta = 0.7$ are used in our numerical simulations. When the reference distance is $1$~m, the transmit power is configured as $30$~dBm, the path loss is set as $-20$~dB, and the Rician factor is $3$~dB.  The Gaussian randomization process utilizes $1000$ candidate random vectors.  The convergence criterion for the proposed algorithm is set to $10^{-3}$. The number of transmit antennas is $N_{T}=64$.

\begin{figure}[htbp]
\centering
\begin{minipage}[t]{0.48\textwidth}
\centering
\includegraphics[scale=0.55]{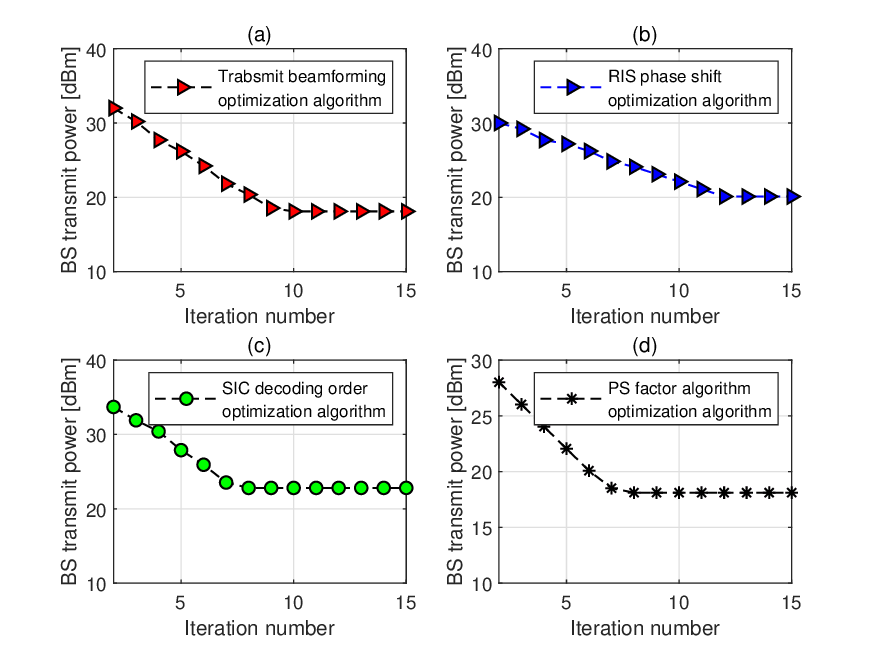}
\caption{Transmit power versus the number of iterations.}
\label{FIGURE2}
\end{minipage}
\begin{minipage}[t]{0.48\textwidth}
\centering
\includegraphics[scale=0.55]{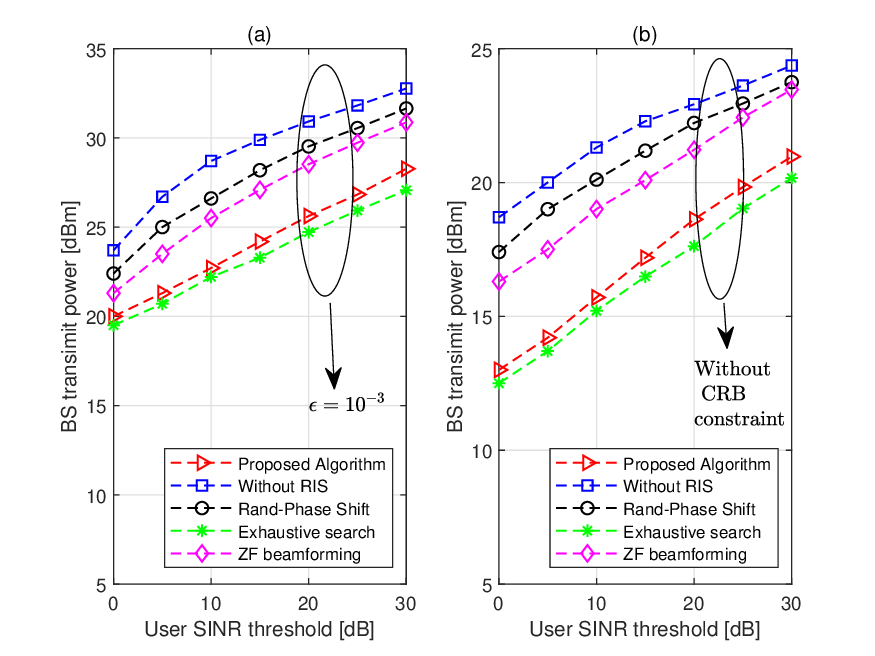}
\caption{Power consumption versus SINR requirement of the IoT device .}
\label{FIGURE3}
\end{minipage}
\end{figure}
In Fig.~\ref{FIGURE2}, the exhaustive search method is used to obtain the optimal SIC decoding order.
Fig.~\ref{FIGURE2} illustrates the convergence trend of each sub-algorithm introduced in this section. This paper studies the convergence performance of the four sub-algorithms within the proposed algorithm. It shows that power consumption reduces as the number of iterations increases in each sub-algorithm. This simulation shows that all sub-algorithms converge, and the convergence occurs between the $6$-th and $12$-th iterations. Finally, since the proposed sub-algorithm converges, the proposed BCD algorithm is convergence.

\begin{figure}[htbp]
\centering
\begin{minipage}[t]{0.48\textwidth}
\centering
\includegraphics[scale=0.55]{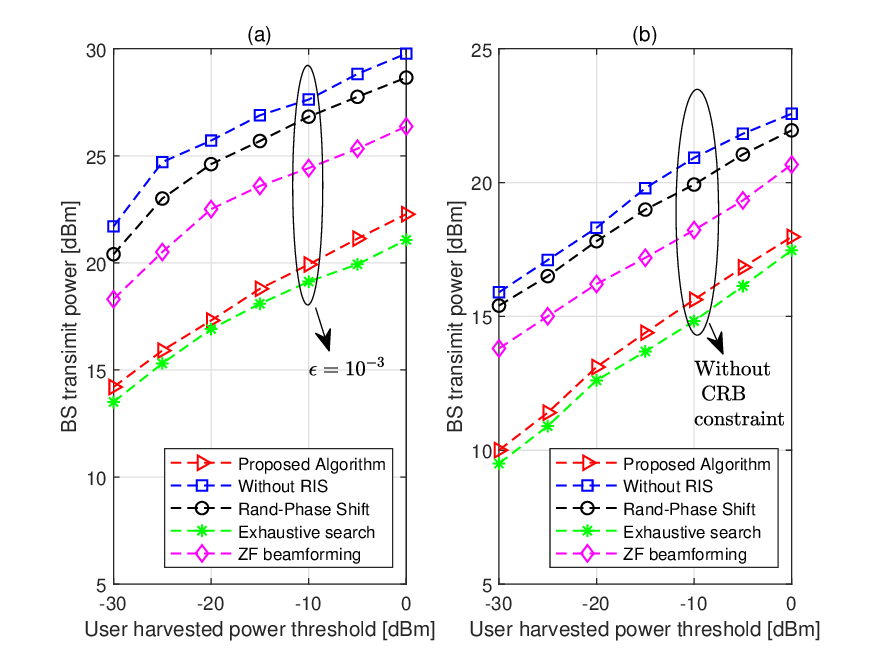}
\caption{Power consumption versus IoT device energy harvesting threshold.}
\label{FIGURE4}
\end{minipage}
\begin{minipage}[t]{0.48\textwidth}
\centering
\includegraphics[scale=0.55]{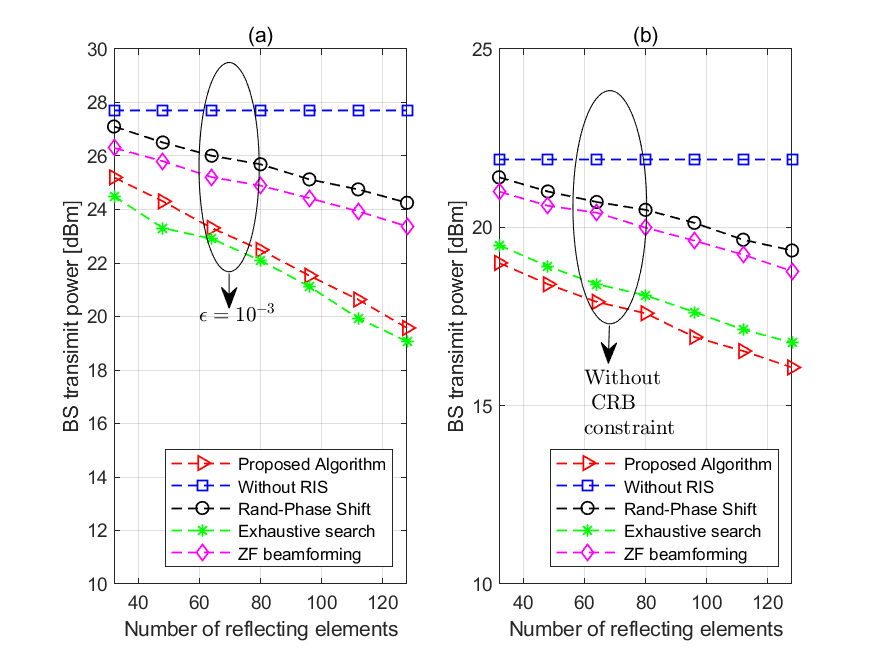}
\caption{Power consumption versus number of RIS reflecting elements}
\label{FIGURE5}
\end{minipage}
\end{figure}

Subsequently, we compare the proposed algorithm with four baseline algorithms. The RIS is not considered during operation in the baseline algorithm \textbf{Without RIS}, while the optimization process for other variables aligns with the proposed algorithm framework. Baseline algorithm \textbf{Exhaustive search} employs an exhaustive search method to optimize the SIC decoding order, maintaining consistency with the proposed algorithm in all other aspects. Similarly, baseline algorithm \textbf{zero-forcing (ZF) beamforming} uses the ZF method to optimize transmit beamforming, with the rest of the sub-algorithms being consistent with the proposed algorithm. Additionally, baseline algorithm \textbf{Rand-phase shift} applies a random optimization algorithm to adjust the RIS phase shift,  and other algorithms remain the same as those in the proposed algorithm. Among these baseline algorithms, algorithm C is identified as the optimal algorithm, and its performance is considered the upper bound. Fig.~\ref{FIGURE3} and Fig.~\ref{FIGURE4} show that the BS transmit power increases as IoT devices' SINR threshold and energy-harvested constraints rise. This increase in power consumption is due to the need for more power to meet the higher SINR and energy-harvested threshold. Furthermore, the simulation results demonstrate that the proposed algorithm's performance is closer to the exhaustive search algorithm while significantly reducing computational complexity. Moreover, the proposed algorithm outperforms the other baseline algorithms. As sensing accuracy increases, the power requirement also rises because the BS needs more power to satisfy the sensing accuracy constraint.

The variation of BS transmit power with the number of RIS reflecting elements under different algorithms is depicted in Fig.~\ref{FIGURE5}. The trend shows that increasing the number of RIS-reflecting elements results in a continuous decrease in BS transmit power. This is attributed to the ability of the RIS to adjust the channel, enhancing system performance and gradually reducing BS transmit power. The comparison between ISCPT systems with and without RIS in Fig.~ \ref{FIGURE5} reveals that RIS improves system performance. When considering the sensing accuracy constraint of $\epsilon=10^{-3}$ and comparing it with systems without CRB constraint, it is evident that RIS plays a crucial role in meeting sensing requirements in ISCPT NOMA systems. Our proposed algorithm outperforms others mainly due to its ability to achieve global convergence, unlike the others.

\begin{figure}[htbp]
\centering
\includegraphics[scale=0.3]{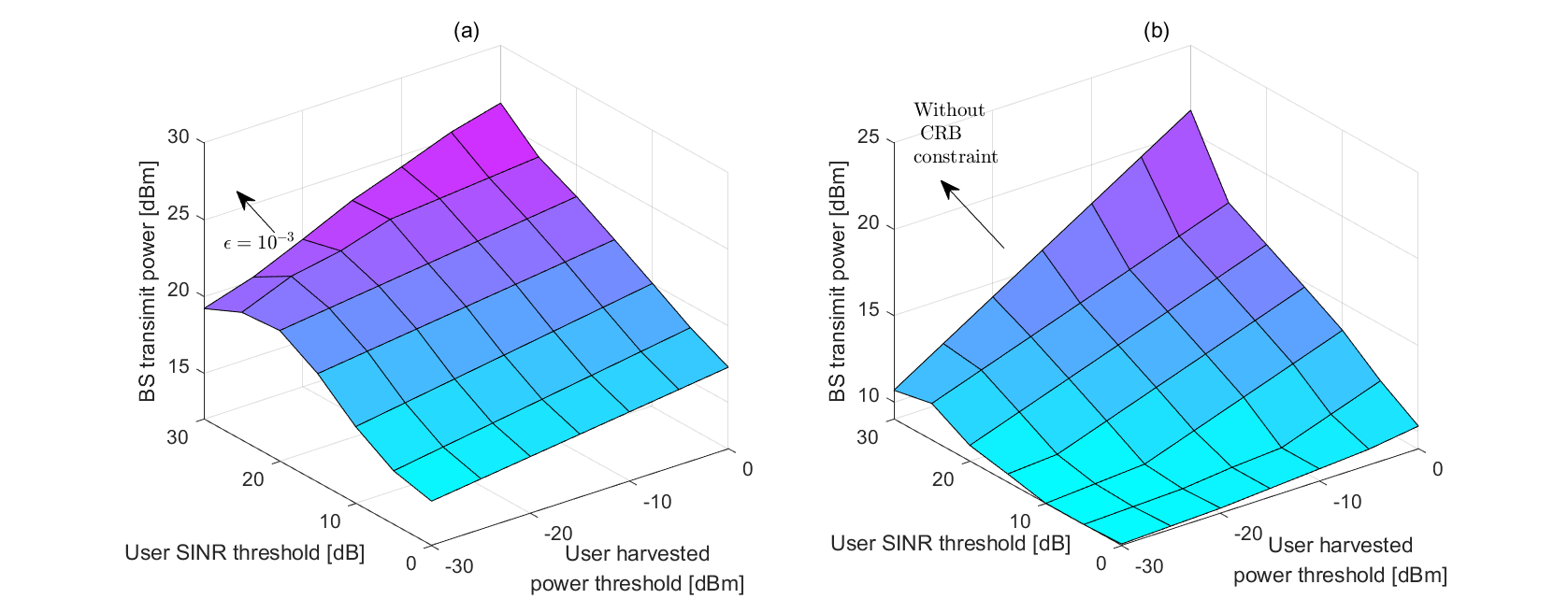}
\caption{Transmit Power versus SINR and IoT device energy harvested thresholds with (a) sensing CRB $\epsilon=10^{-3}$ and (b) without CRB constraint.}
\label{FIGURE6}
\end{figure}

The performance of target positioning based on ISCPT is illustrated in Figure .\ref{FIGURE6}. The target is positioned within a circle defined by parameters $30m$ and $50m$. It can be observed that the proposed algorithm saves more BS transmit power while satisfying the sensing accuracy, IoT devices' SINR, and energy harvesting constraints. As expected, more BS transmit power is required as the distance increases. This is due to the increased path loss, which results in a smaller target reflection energy, necessitating more BS transmit power to compensate for the sensing accuracy loss. The simulation results also demonstrate that while meeting the basic requirements of communication and power transfer, the position of the target significantly influences BS transmit power. Finally, when the SINR threshold or the energy harvesting threshold is higher, target positioning has a substantial impact on BS transmit power consumption, highlighting the importance of balancing sensing, communication, and power transfer.

\begin{figure}[htbp]
\centering
\includegraphics[scale=0.55]{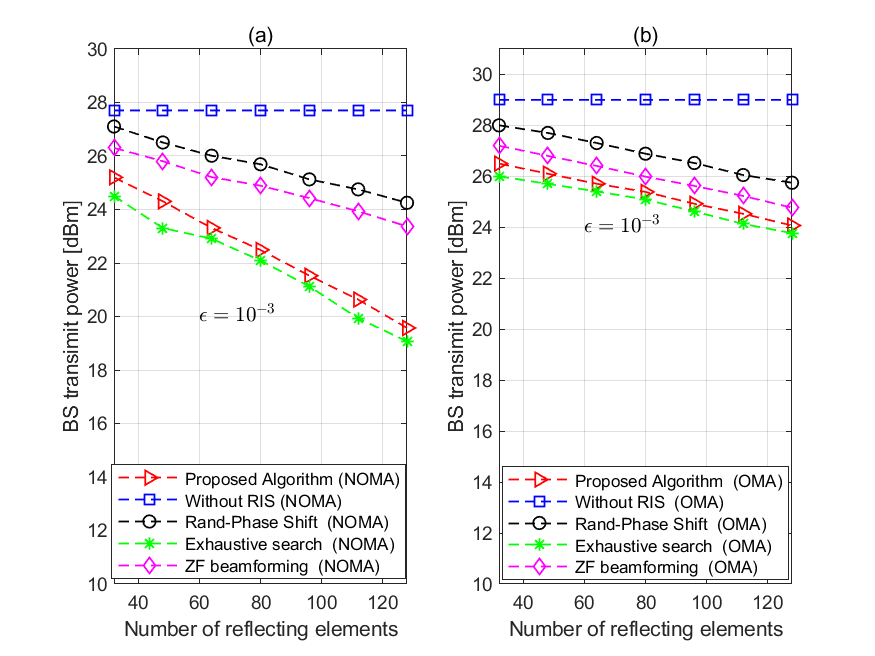}
\caption{Transmit Power with NOMA versus the transmit Power with OMA in (a) sensing CRB $\epsilon=10^{-3}$ and (b) sensing CRB $\epsilon=10^{-3}$.}
\label{FIGURE7}
\end{figure}

As shown in Fig~\ref{FIGURE7}, NOMA is less power consumption than OMA, this is because it allows multiple users to share the same spectrum simultaneously by using power-domain separation, enabling better spectral efficiency. NOMA optimizes power by allocating more power to users with poor channel conditions and less power to users with strong channels, while SIC helps manage interference. This results in lower total power consumption compared to OMA, which assigns separate resources to users and cannot exploit these efficiencies.

\begin{figure}[htbp]
\begin{minipage}[t]{0.48\textwidth}
\centering
\includegraphics[scale=0.55]{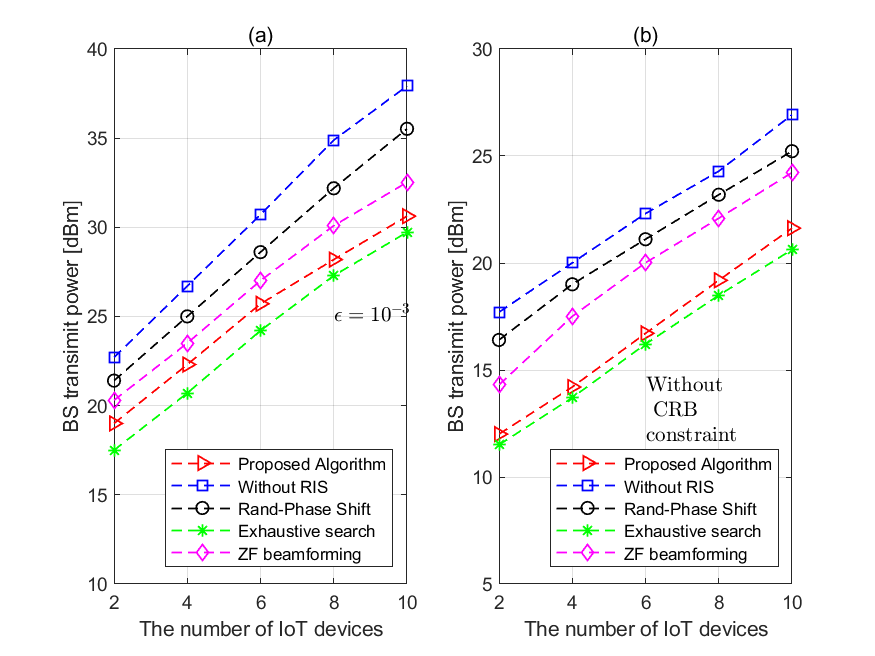}
\caption{Power consumption versus the number of IoT devices .}
\label{FIGURE8}
\end{minipage}
\end{figure}

As shown in Fig~\ref{FIGURE8}, we observed a significant increase in the ISCPT's power consumption. In addition, all algorithms employed in this context are experiencing a rise in energy consumption as the number of devices continues to grow. This is because the power consumption is primarily attributed to the need to satisfy the more sensing and communication constraints of these devices, which collectively contribute to increased power consumption. In Fig~\ref{FIGURE8}, despite the inherent complexity and demands of exhaustive search algorithms, our suggested approach closely approximates their performance while consuming substantially less energy. The findings show the proposed optimization algorithm's capacity to maintain stable power consumption in the face of the rising number of IoT devices.

\section{Conclusion and Future Works}\label{VI}
This paper focuses on minimizing the transmit power of the RIS-assisted ISCPT-NOMA system. We optimize the SIC decoding sequence, transmit beamforming, RIS phase shift, and PS scale factor to achieve power minimization under IoT device QoS, sensing accuracy CRB, energy harvesting, and modulus value of RIS reflection units. The interdependence of these variables makes the problem non-convex. To address this, the original problem is divided into four non-convex sub-problems. Then, we proposed a BCD algorithm based on SCA, SDR, and ADMM to solve each sub-problem. We analyze the convergence and computational complexity of the proposed BCD algorithm. Simulation results demonstrate that the proposed algorithm effectively reduces BS transmit power while balancing sensing accuracy and energy harvesting. Future research will focus on the system's robustness, considering the imperfections in the system's channel state information.


\appendices
\section{Proof of the Expression in (\ref{pro13})}\label{appA}
According to the theory of matrix derivatives in \cite{b38}, the derivatives of $\boldsymbol{\eta}$ is expressed as
\begin{small}
\begin{align}
&\frac{\partial\boldsymbol{\eta}}{\partial\theta}=\alpha_{t}\mathrm{vec}\left(\frac{\partial{\mathbf{G}^{H}\boldsymbol{\Theta}^{H}\mathbf{h}_{s}\mathbf{h}_{s}^{H}\boldsymbol{\Theta}\mathbf{G}}}{\partial\theta}\mathbf{W}\mathbf{S}\right)
=\alpha_{t}\mathrm{vec}\left(\left((\mathbf{G}^{H}\boldsymbol{\Theta}^{H}\right.\right.\nonumber\\
&(\mathbf{h}_{s}\odot\mathbf{e})\mathbf{h}_{s}^{H}\boldsymbol{\Theta}\mathbf{G})\left.\left.+(\mathbf{G}^{H}\boldsymbol{\Theta}^{H}\mathbf{h}_{s}(\mathbf{h}_{s}\odot\mathbf{e})^{H}\boldsymbol{\Theta}\mathbf{G})\right)\mathbf{W}\mathbf{S}\right),\label{proA1_1}
\end{align}
\end{small}%
where $\mathbf{e}=\left[0,-j\pi\cos\theta,\ldots,-j(M-1)\pi\cos\theta\right]^{T}\in\mathbb{C}^{M\times 1}$.
Similarly, the derivatives of $\boldsymbol{\eta}$ with respect to $\boldsymbol{\alpha}$ is given by
\begin{small}
\begin{align}
\frac{\partial\boldsymbol{\eta}}{\partial\boldsymbol{\alpha}}=[1,j]^{T}\otimes\mathrm{vec}\left(\mathbf{G}^{H}\boldsymbol{\Theta}^{H}\mathbf{h}_{s}\mathbf{h}_{s}^{H}\boldsymbol{\Theta}\mathbf{G}\mathbf{W}\mathbf{S}\right).\label{proA1_2}
\end{align}
\end{small}%
Based on (\ref{proA1_1}) and (\ref{proA1_2}), the second order partial derivatives of $\mathbf{F}_{\theta,\theta}$, $\mathbf{F}_{\boldsymbol{\alpha},\boldsymbol{\alpha}^{T}}$ and $\mathbf{F}_{\theta,\boldsymbol{\alpha}^{T}}$ are given by
\begin{small}
\begin{align}
&\mathbf{F}_{\theta,\theta}=\frac{2|\alpha|^{2}}{\sigma_{r}^{2}}\mathrm{Re}(\mathrm{vec}^{H}(((\mathbf{G}^{H}\boldsymbol{\Theta}^{H}(\mathbf{h}_{s}\odot\mathbf{e})\mathbf{h}_{s}^{H}\boldsymbol{\Theta}\mathbf{G})+(\mathbf{G}^{H}\boldsymbol{\Theta}^{H}\mathbf{h}_{s}\nonumber\\
&(\mathbf{h}_{s}\odot\mathbf{e})^{H}\boldsymbol{\Theta}\mathbf{G}))\mathbf{W}\mathbf{S})\mathrm{vec}(((\mathbf{G}^{H}\boldsymbol{\Theta}^{H}(\mathbf{h}_{s}\odot\mathbf{e})\mathbf{h}_{s}^{H}\boldsymbol{\Theta}\mathbf{G})+(\mathbf{G}^{H}\boldsymbol{\Theta}^{H}\nonumber\\
&\mathbf{h}_{s}(\mathbf{h}_{s}\odot\mathbf{e})^{H}\boldsymbol{\Theta}\mathbf{G}))\mathbf{W}\mathbf{S}))=\frac{2L|\alpha|^{2}}{\sigma_{r}^{2}}
\mathrm{tr}(((\mathbf{G}^{H}\boldsymbol{\Theta}^{H}(\mathbf{h}_{s}\odot\mathbf{e})\mathbf{h}_{s}^{H}\boldsymbol{\Theta}\mathbf{G})\nonumber\\
&+(\mathbf{G}^{H}\boldsymbol{\Theta}^{H}\mathbf{h}_{s}(\mathbf{h}_{s}\odot\mathbf{e})^{H}\boldsymbol{\Theta}\mathbf{G}))\mathbf{W}\mathbf{W}^{H}((\mathbf{G}^{H}\boldsymbol{\Theta}^{H}(\mathbf{h}_{s}\odot\mathbf{e})\mathbf{h}_{s}^{H}\boldsymbol{\Theta}\mathbf{G})\nonumber\\
&+(\mathbf{G}^{H}\boldsymbol{\Theta}^{H}\mathbf{h}_{s}(\mathbf{h}_{s}\odot\mathbf{e})^{H}\boldsymbol{\Theta}\mathbf{G}))^{H}),\label{proA1_3}
\end{align}
\end{small}%
and
\begin{small}
\begin{align}
&\mathbf{F}_{\boldsymbol{\alpha},\boldsymbol{\alpha}^{T}}=\frac{2L}{\sigma_{r}^{2}}\mathrm{tr}(\mathbf{G}^{H}\boldsymbol{\Theta}^{H}\mathbf{h}_{s}\mathbf{h}_{s}^{H}\boldsymbol{\Theta}\mathbf{G}\mathbf{W}\mathbf{W}^{H}(\mathbf{G}^{H}\boldsymbol{\Theta}^{H}\mathbf{h}_{s}\mathbf{h}_{s}^{H}\boldsymbol{\Theta}\mathbf{G})^{H})\nonumber\\
&\mathbf{I}_{2},\label{proA1_4}
\end{align}
\end{small}%
and
\begin{small}
\begin{align}
&\mathbf{F}_{\theta,\boldsymbol{\alpha}^{T}}=\frac{2L}{\sigma_{r}^{2}}\mathrm{Re}(\mathrm{tr}(\mathbf{G}^{H}\boldsymbol{\Theta}^{H}\mathbf{h}_{s}\mathbf{h}_{s}^{H}\boldsymbol{\Theta}\mathbf{G}\mathbf{W}\mathbf{W}^{H}((\mathbf{G}^{H}\boldsymbol{\Theta}^{H}(\mathbf{h}_{s}\odot\mathbf{e})\nonumber\\
&\mathbf{h}_{s}^{H}\boldsymbol{\Theta}\mathbf{G})+(\mathbf{G}^{H}\boldsymbol{\Theta}^{H}\mathbf{h}_{s}(\mathbf{h}_{s}\odot\mathbf{e})^{H}\boldsymbol{\Theta}\mathbf{G}))^{H})[1,j]).\label{proA1_5}
\end{align}
\end{small}%
The proof of expression of $\mathbf{F}_{\theta,\theta}$, $\mathbf{F}_{\boldsymbol{\alpha},\boldsymbol{\alpha}^{T}}$ and $\mathbf{F}_{\theta,\boldsymbol{\alpha}^{T}}$ are completed.

\section{Proof of Rank one}\label{appB}
According to \cite{b35}, we find that the transmit beamforming optimization problem is convex and satisfies Slater's condition. Therefore, the duality gap is $0$.  The optimal solution is obtained by dealing with the dual problem. Let 
$\mathbf{W}_{k}=\mathbf{w}_{k}\mathbf{w}_{k}^{H},~\forall~k$, the Lagrangian function of problem is given by
\begin{small}
\begin{align}
&\mathcal{L}=\sum_{i=1}^{K}\mathrm{tr}(\mathbf{W}_{k})+\lambda\boldsymbol{u}^{T}\left[\begin{matrix}
\zeta_{1}&\zeta_{2}\\
\zeta_{3}^{T}&\zeta_{3}\\
  \end{matrix}
  \right]\boldsymbol{u}^{T}+\bar{\lambda}(\frac{1}{\zeta}-\epsilon)+\lambda_{1}(\sum\nolimits_{i=1}^{K}\nonumber\\
&\mathrm{tr}(\tilde{\mathbf{H}}_{t}(\phi)\mathbf{W}_{k}\tilde{\mathbf{H}}_{t}^{H}(\phi))-\frac{\sigma_{r}^{2}(\zeta_{1}+\zeta)}{2L|\alpha_{t}|^{2}})+\lambda_{2}(\sum_{i=1}^{K}\mathrm{tr}(\mathbf{H}_{t}(\phi)\mathbf{W}_{k}\mathbf{H}_{t}^{H}(\phi))\nonumber\\
&\mathbf{I}_{2}-\frac{\sigma_{r}^{2}\zeta_{2}}{2L})+\lambda_{3}(\sum\nolimits_{i=1}^{K}\mathrm{Re}(\mathrm{tr}(\mathbf{H}_{t}(\phi)\mathbf{W}_{k}\tilde{\mathbf{H}}_{t}^{H}(\phi))[1,j]) -\frac{\sigma_{r}^{2}\zeta_{3}}{2L})+\nonumber\\
&\lambda_{4}\sum_{i=1}^{K}(\rho_{k}\mathbf{h}^{H}_{u,i}\boldsymbol{\Theta}\mathbf{G}\mathbf{W}_{k}\mathbf{G}^{H}\boldsymbol{\Theta}^{H}\mathbf{h}_{u,i}-\rho_{k}\sum_{d(j)\leq d(k)}\mathbf{h}^{H}_{u,i}\boldsymbol{\Theta}\mathbf{G}\mathbf{W}_{j}\mathbf{G}^{H}\boldsymbol{\Theta}^{H}\mathbf{h}_{u,i}\nonumber\\
&-\rho_{k}\sigma_{k}^{2}-\delta_{k}^{2}(2^{\gamma_{k}}-1))+\lambda_{5}\sum_{i=1}^{K}(\mathbf{G}^{T}\boldsymbol{\Theta}\mathbf{h}_{r,i}\mathbf{W}_{k}\mathbf{h}_{r,i}^{H}\boldsymbol{\Theta}^{H}\mathbf{G}^{*}-\nonumber\\
&(2^{\gamma_{k}}-1)\sum\nolimits_{j\neq i}\mathbf{G}^{T}\boldsymbol{\Theta}\mathbf{h}_{r,i}\mathbf{W}_{j}\mathbf{h}_{r,i}^{H}\boldsymbol{\Theta}^{H}\mathbf{G}^{*}+\sigma_{k}^{2})+\lambda_{6}\sum_{i=1}^{K}(\bar{f}_{1}(\mathbf{W}_{k})-\nonumber\\
&\ln\left(\sum_{d(j)>d(k)}\mathbf{h}^{H}_{u,i}\boldsymbol{\Theta}\mathbf{G}\mathbf{W}_{j}\mathbf{G}^{H}\boldsymbol{\Theta}^{H}\mathbf{h}_{u,i}+A_{k}\right)-\nonumber\\
&\ln(\mathbf{h}^{H}_{u,i}\boldsymbol{\Theta}\mathbf{G}\mathbf{W}_{k}\mathbf{G}^{H}\boldsymbol{\Theta}^{H}\mathbf{h}_{u,\bar{k}})+\bar{f}_{2}(\mathbf{W}_{j})+\lambda_{7}\sum_{i=1}^{K}(\eta_{k}(1-\rho_{k})(\mathbf{h}^{H}_{u,i}\boldsymbol{\Theta}\mathbf{G}\nonumber\\
&\mathbf{W}_{k}\mathbf{G}^{H}\boldsymbol{\Theta}^{H}\mathbf{h}_{u,i}+\sigma_{k}^{2})-Q_{k})-\sum\nolimits_{i=1}^{K}\mathrm{tr}(\mathbf{W}_{k}\mathbf{Y}_{k}).\label{proA2_1}
\end{align}
\end{small}
where $\tilde{\mathbf{H}}_{t}=((\mathbf{G}^{H}\boldsymbol{\Theta}^{H}(\mathbf{h}_{s}\odot\mathbf{e})\mathbf{h}_{s}^{H}\boldsymbol{\Theta}\mathbf{G})+(\mathbf{G}^{H}\boldsymbol{\Theta}^{H}\mathbf{h}_{s}(\mathbf{h}_{s}\odot\mathbf{e})^{H}$\\
$\boldsymbol{\Theta}\mathbf{G}))$, $\mathbf{H}_{t}=\mathbf{G}^{H}\boldsymbol{\Theta}^{H}\mathbf{h}_{s}\mathbf{h}_{s}^{H}\boldsymbol{\Theta}\mathbf{G}$.
$\mathbf{Y}_{k}\in\mathbb{C}^{N_{T}\times N_{T}}$ is the Lagrangian multiplier matrix. The dual problem of the problem is
\begin{small}
\begin{align}
\max_{\lambda_{k},\mu_{k\bar{k}},v_{\bar{k}},\varpi_{k}=0, \mathbf{Y}_{k}\succeq 0}\min_{\mathbf{W}_{k},\varphi_{k},\phi_{\bar{k}}}\mathcal{L}
\end{align}
\end{small}%
Next, we apply the Karush-Kuhn-Tucker (KKT) conditions to investigate the optimal solution structure of the dual problem. The KKT condition related to $\mathbf{W}_{k}^{*}$ can be expressed as follows
\begin{small}
\begin{align}
K_{1}:\lambda^{*},\bar{\lambda}^{*},\lambda_{1}^{*},\lambda_{2}^{*},\lambda_{3}^{*},\lambda_{4}^{*},\lambda_{5}^{*},\lambda_{6}^{*},\lambda_{7}^{*}>0,\mathbf{Y}_{k}^{*}\succeq 0\\
K_{2}: \mathbf{W}_{k}^{*}\mathbf{Y}_{k}^{*}=0\\
K_{3}: \nabla_{\mathbf{W}_{k}^{*}}\mathcal{L}=0\label{poA2__3}
\end{align}
\end{small}%
where $\lambda^{*},\bar{\lambda}^{*},\lambda_{1}^{*},\lambda_{2}^{*},\lambda_{3}^{*},\lambda_{4}^{*},\lambda_{5}^{*},\lambda_{6}^{*},\lambda_{7}^{*}$ and $\mathbf{Y}_{k}^{*}$ represent the optimal Lagrangian multiples of the dual problem. $\nabla_{\mathbf{W}_{k}^{*}}\mathcal{L}$ denotes the gradient vector of Eq.(\ref{proA2_1}) w.r.t $\mathbf{W}_{k}^{*}$. We can explicitly express $K_{3}$ as follows
\begin{small}
\begin{align}
\mathbf{Y}_{k}^{*}=\mathbf{I}-\boldsymbol{\Delta},\label{poA2_4}
\end{align}
\end{small}%
where $\boldsymbol{\Delta}$ can be given by
\begin{small}
\begin{align}
&\boldsymbol{\Delta}=\lambda_{1}\tilde{\boldsymbol{\Gamma}}_{t}+\lambda_{2}\bar{\boldsymbol{\Gamma}}_{t}+\lambda_{3}\hat{\boldsymbol{\Gamma}}_{t}+\lambda_{4}\boldsymbol{C}_{k}+\lambda_{5}\boldsymbol{J}_{k}+\lambda_{6}\boldsymbol{U}_{k}+\lambda_{7}\boldsymbol{V}_{k},\label{poA2_5}
\end{align}
\end{small}%
where $\tilde{\boldsymbol{\Gamma}}_{t}=\tilde{\mathbf{H}}_{t}^{H}(\phi)\tilde{\mathbf{H}}_{t}(\phi)$, $\bar{\boldsymbol{\Gamma}}_{t}=\mathbf{H}_{t}^{H}(\phi)\mathbf{H}_{t}(\phi)$, $\hat{\boldsymbol{\Gamma}}_{t}=\tilde{\mathbf{H}}_{t}^{H}(\phi)\mathbf{H}_{t}(\phi)$, $\boldsymbol{C}_{k}=\rho_{k}\mathbf{G}^{H}\boldsymbol{\Theta}^{H}\mathbf{h}_{u,i}\mathbf{h}^{H}_{u,i}\boldsymbol{\Theta}\mathbf{G}$,$\boldsymbol{J}_{k}=\mathbf{G}^{T}\boldsymbol{\Theta}\mathbf{h}_{r,i}\mathbf{W}_{k}\mathbf{h}_{r,i}^{H}\boldsymbol{\Theta}^{H}\mathbf{G}^{*}$, $\boldsymbol{U}_{k}=\frac{\mathbf{G}^{H}\boldsymbol{\Theta}^{H}\mathbf{h}_{u,\bar{k}}\mathbf{h}^{H}_{u,i}\boldsymbol{\Theta}\mathbf{G}}{\mathbf{h}^{H}_{u,i}\boldsymbol{\Theta}\mathbf{G}\mathbf{W}_{k}^{*}\mathbf{G}^{H}\boldsymbol{\Theta}^{H}\mathbf{h}_{u,\bar{k}}}$, $\boldsymbol{V}_{k}=\eta_{k}(1-\rho_{k})\mathbf{G}^{H}\boldsymbol{\Theta}^{H}\mathbf{h}_{u,i}\mathbf{h}^{H}_{u,i}\boldsymbol{\Theta}\mathbf{G}$.
Subsequently, we will demonstrate the rank-one property of the beamforming matrix $\mathbf{W}_{k}^{*}$ by examining the structure of 
$\mathbf{Y}_{k}^{*}$. We assign the maximum eigenvalue of 
$\boldsymbol{\Delta}$ as $\zeta_{max}$. It should be noted that, given the channel’s randomness, the likelihood of multiple eigenvalues sharing the same maximum value is negligible. If $\zeta_{max}>1$,  $\mathbf{Y}_{k}^{*}$ cannot be positive semidefinite, contradicting 
$K_{1}$. In the case of $\zeta_{max}<1$, $\mathbf{Y}_{k}^{*}$ must be positive, definite, and full rank. As indicated by 
$K_{2}$, $\mathbf{W}_{k}^{*}$ can only be $\mathbf{0}$, which is clearly inconsistent with reality. Thus, it is imperative that 
$\zeta{max}<1$, leading to $\mathrm{Rank}(\mathbf{Y}_{k}^{*})=N_{T}-1$. Consequently, $\mathrm{Rank}(\mathbf{W}_{k}^{*})=1$, indicating that the beamforming matrix $\mathbf{W}_{k}^{*}$ is rank-one. This concludes the proof.

\section{Proof of formulation (\ref{pro27}) and (\ref{pro28})}\label{appC}
According to (\ref{pro18}) and (\ref{pro21}), These formulations in (\ref{pro18}) can be rewritten as
\begin{align}
&\text{The first formulation of the left hand of (\ref{pro18})}=\bar{g}_{1}(\boldsymbol{\theta})+g_{1}\nonumber\\
&(\boldsymbol{\theta})+g_{2}(\boldsymbol{\theta})+g_{3}(\boldsymbol{\theta})+g_{4}(\boldsymbol{\theta})=\circled{1}\times\circled{3}+\circled{1}\times\circled{4}+\circled{2}\nonumber\\
&\times\circled{3}+\circled{2}\times\circled{4},\nonumber\\
&\text{The second formulation of the left hand of (\ref{pro18})}=\bar{g}_{4}(\boldsymbol{\theta})=\nonumber\\
&\circled{5}\times\circled{5}^{H},\nonumber\\
&\text{The third formulation of the left hand of (\ref{pro18})}=\nonumber\\
&\left[\begin{matrix}
\bar{g}_{2}(\boldsymbol{\theta})=\circled{5}\times\circled{3}\\
\bar{g}_{3}(\boldsymbol{\theta})=\circled{5}\times\circled{4}  
\end{matrix}
\right],\nonumber\\
&(\ref{pro21})\Rightarrow\frac{\circled{6}}{\circled{9}+\circled{11}}\leq\frac{\circled{8}}{\circled{7}+\circled{10}}\Rightarrow\underbrace{\circled{6}\times\circled{7}}_{g_{5}(\boldsymbol{\theta})}-\underbrace{\circled{8}\times\circled{9}}_{g_{6}(\boldsymbol{\theta})}\nonumber\\
&+\underbrace{\circled{6}\times\circled{10}}_{g_{7}(\boldsymbol{\theta})}-\underbrace{\circled{8}\times\circled{11}}_{g_{8}(\boldsymbol{\theta})}\leq 0, \label{proA3_0}  
\end{align}
where $\circled{1}$, $\circled{2}$, $\circled{3}$, $\circled{4}$, $\circled{5}$, $\circled{6}$, $\circled{7}$, $\circled{8}$, $\circled{9}$, $\circled{10}$, $\circled{11}$ are given by (\ref{proA3_1}) on the top of the next page, where $\boldsymbol{\Omega}_{2}=((\mathrm{diag}((\mathbf{h}_{s}\odot\mathbf{e})^{H})\mathbf{G}\mathbf{G}^{H}\mathrm{diag}(\mathbf{h}_{s}))^{T}\otimes(\mathrm{diag}(\mathbf{w}_{k}^{H}\mathbf{G}^{H})\mathbf{h}_{s}))$, $\bar{\boldsymbol{\Omega}}_{2}^{T}=\mathrm{vec}((\mathbf{h}_{s}\odot\mathbf{e})^{H}\mathrm{diag}(\mathbf{G}\mathbf{w}_{k}))$, $\tilde{\boldsymbol{\Omega}}_{2}^{T}=\bar{\boldsymbol{\Omega}}_{2}^{T}\otimes\boldsymbol{\Omega}_{2}$,
$\boldsymbol{\Omega}_{3}=((\mathrm{diag}(\mathbf{h}_{s}^{H})\mathbf{G}\mathbf{G}^{H}\mathrm{diag}((\mathbf{h}_{s}\odot\mathbf{e})))^{T}\otimes(\mathrm{diag}(\mathbf{w}_{k}^{H}\mathbf{G}^{H})(\mathbf{h}_{s}\odot\mathbf{e})))$, $\bar{\boldsymbol{\Omega}}_{3}=\mathrm{vec}(\mathbf{h}_{s}^{H}\mathrm{diag}(\mathbf{G}\mathbf{w}_{k}))$, $\tilde{\boldsymbol{\Omega}}_{3}^{T}=\bar{\boldsymbol{\Omega}}_{3}^{T}\otimes\boldsymbol{\Omega}_{3}$,
$\boldsymbol{\Omega}_{4}=((\mathrm{diag}(\mathbf{h}_{s}^{H})\mathbf{G}\mathbf{G}^{H}\mathrm{diag}(\mathbf{h}_{s}))^{T}\otimes(\mathrm{diag}(\mathbf{w}_{k}^{H}\mathbf{G}^{H})(\mathbf{h}_{s}\odot\mathbf{e})))$, $\bar{\boldsymbol{\Omega}}_{4}=\mathrm{vec}((\mathbf{h}_{s}\odot\mathbf{e})^{H}\mathrm{diag}(\mathbf{G}\mathbf{w}_{k}))$, $\tilde{\boldsymbol{\Omega}}_{4}^{T}=\bar{\boldsymbol{\Omega}}_{4}^{T}\otimes\boldsymbol{\Omega}_{4}$,
$\boldsymbol{\Omega}_{6}=$\\
$((\mathrm{diag}(\mathbf{h}_{s}^{H})\mathbf{G}\mathbf{G}^{H}\mathrm{diag}((\mathbf{h}_{s})))^{T}\otimes(\mathrm{diag}(\mathbf{w}_{k}^{H}\mathbf{G}^{H})(\mathbf{h}_{s}\odot\mathbf{e}))$\\
$)$,$\bar{\boldsymbol{\Omega}}_{6}=\mathrm{vec}(\mathbf{h}_{s}^{H}\mathrm{diag}(\mathbf{G}\mathbf{w}_{k}))$, $\tilde{\boldsymbol{\Omega}}_{6}^{T}=\bar{\boldsymbol{\Omega}}_{6}^{T}\otimes\boldsymbol{\Omega}_{6}$,
$((\mathrm{diag}(\mathbf{h}_{s}^{H})\mathbf{G}\mathbf{G}^{H}\mathrm{diag}(\mathbf{h}_{s}))^{T}\otimes(\mathrm{diag}(\mathbf{w}_{k}^{H}\mathbf{G}^{H})(\mathbf{h}_{s})))=\boldsymbol{\Omega}_{7}$,$\mathrm{vec}((\mathbf{h}_{s}\odot\mathbf{e})^{H}\mathrm{diag}(\mathbf{G}\mathbf{w}_{k}))=\bar{\boldsymbol{\Omega}}_{7}$, $\tilde{\boldsymbol{\Omega}}_{7}^{T}=\bar{\boldsymbol{\Omega}}_{7}^{T}\otimes\boldsymbol{\Omega}_{7}$, $\tilde{\boldsymbol{\Omega}}_{7}^{T}=\bar{\boldsymbol{\Omega}}_{7}^{T}\otimes\boldsymbol{\Omega}_{7}$,
$\boldsymbol{\Omega}_{8}=((\mathrm{diag}(\mathbf{h}_{s}^{H})\mathbf{G}\mathbf{G}^{H}\mathrm{diag}(\mathbf{h}_{s}))^{T}\otimes(\mathrm{diag}(\mathbf{w}_{k}^{H}\mathbf{G}^{H})(\mathbf{h}_{s})))$, $\bar{\boldsymbol{\Omega}}_{8}=\mathrm{vec}((\mathbf{h}_{s})^{H}\mathrm{diag}(\mathbf{G}\mathbf{w}_{k}))$, $\tilde{\boldsymbol{\Omega}}_{8}^{T}=\bar{\boldsymbol{\Omega}}_{8}^{T}\otimes\boldsymbol{\Omega}_{8}$. we use the vectorization operation in\cite{b38}, $g_{1}(\boldsymbol{\theta})$, $g_{2}(\boldsymbol{\theta})$, $g_{3}(\boldsymbol{\theta})$, $g_{4}(\boldsymbol{\theta})$, $\bar{g}_{2}(\boldsymbol{\theta})$, $\bar{g}_{3}(\boldsymbol{\theta})$, $\bar{g}_{4}(\boldsymbol{\theta})$ are given by (\ref{proA3__1}) on the top of the next page.
\begin{figure*}[htbp] 
\centering 
\vspace*{6pt} 
\begin{small}
\begin{align}
&\circled{1}=\mathbf{w}_{k}^{H}\mathbf{G}^{H}\boldsymbol{\Theta}^{H}\mathbf{h}_{s}(\mathbf{h}_{s}\odot\mathbf{e})^{H}\boldsymbol{\Theta}\mathbf{G},
\circled{2}=\mathbf{w}_{k}^{H}\mathbf{G}^{H}\boldsymbol{\Theta}^{H}(\mathbf{h}_{s}\odot\mathbf{e})^{H}\mathbf{h}_{s}\boldsymbol{\Theta}\mathbf{G},\nonumber\\
&\circled{3}=\mathbf{G}^{H}\boldsymbol{\Theta}^{H}(\mathbf{h}_{s}\odot\mathbf{e})\mathbf{h}_{s}^{H}\boldsymbol{\Theta}\mathbf{G}\mathbf{w}_{k},\circled{4}=\mathbf{G}^{H}\boldsymbol{\Theta}^{H}\mathbf{h}_{s}(\mathbf{h}_{s}\odot\mathbf{e})^{H}\boldsymbol{\Theta}\mathbf{G}\mathbf{w}_{k},\nonumber\\
&\circled{5}=\mathbf{w}_{k}^{H}\mathbf{G}^{H}\boldsymbol{\Theta}^{H}\mathbf{h}_{s}\mathbf{h}_{s}^{H}\boldsymbol{\Theta}\mathbf{G}, \circled{6}=\rho_{k}\mathbf{w}_{k}^{H}\mathbf{G}^{H}\boldsymbol{\Theta}^{H}\mathbf{h}_{u,i}\mathbf{h}_{u,i}^{H}\boldsymbol{\Theta}\mathbf{G}\mathbf{w}_{k}, \nonumber\\
&\circled{7}=\sum_{d(j)\leq d(k)}\rho_{\bar{k}}\mathbf{w}_{j}^{H}\mathbf{G}^{H}\boldsymbol{\Theta}^{H}\mathbf{h}_{u,\bar{k}}\mathbf{h}_{u,\bar{k}}^{H}\boldsymbol{\Theta}\mathbf{G}\mathbf{w}_{j}, \circled{8}=\rho_{\bar{k}}\mathbf{w}_{k}^{H}\mathbf{G}^{H}\boldsymbol{\Theta}^{H}\mathbf{h}_{u,\bar{k}}\mathbf{h}_{u,\bar{k}}^{H}\boldsymbol{\Theta}\mathbf{G}\mathbf{w}_{k}, \nonumber\\
&\circled{9}=\sum_{d(j)\leq d(k)}\rho_{\bar{k}}\mathbf{w}_{j}^{H}\mathbf{G}^{H}\boldsymbol{\Theta}^{H}\mathbf{h}_{u,\i}\mathbf{h}_{u,i}^{H}\boldsymbol{\Theta}\mathbf{G}\mathbf{w}_{j}, \circled{10}=\rho_{\bar{k}}\sigma_{\bar{k}}^{2}+\delta_{\bar{k}}^{2},\circled{11}=\rho_{k}\sigma_{k}^{2}+\delta_{k}^{2}. \label{proA3_1}
\end{align}
\end{small}
\hrulefill
\end{figure*}
\begin{figure*}[htbp] 
\centering 
\vspace*{6pt} 
\begin{align}
&g_{1}(\boldsymbol{\theta})=\circled{1}\times\circled{3}=\mathrm{vec}(\boldsymbol{\theta}^{H}\mathrm{diag}(\mathbf{w}_{k}^{H}\mathbf{G}^{H})\mathbf{h}_{s}\boldsymbol{\theta}\mathrm{diag}((\mathbf{h}_{s}\odot\mathbf{e})^{H})\mathbf{G}\mathbf{G}^{H}\mathrm{diag}((\mathbf{h}_{s}\odot\mathbf{e}))\boldsymbol{\theta}^{H})\mathbf{h}_{s}^{H}\boldsymbol{\Theta}\mathbf{G}\mathbf{w}_{k}\nonumber\\
&=(\boldsymbol{\theta}\otimes\boldsymbol{\theta})^{H}\mathrm{vec}(\mathrm{diag}(\mathbf{w}_{k}^{H}\mathbf{G}^{H})\mathbf{h}_{s}\boldsymbol{\theta}\mathrm{diag}((\mathbf{h}_{s}\odot\mathbf{e})^{H})\mathbf{G}\mathbf{G}^{H}\mathrm{diag}((\mathbf{h}_{s}\odot\mathbf{e})))\mathbf{h}_{s}^{H}\mathrm{diag}(\mathbf{G}\mathbf{w}_{k})\boldsymbol{\theta}\nonumber\\
&=(\boldsymbol{\theta}\otimes\boldsymbol{\theta})^{H}\underbrace{((\mathrm{diag}((\mathbf{h}_{s}\odot\mathbf{e})^{H})\mathbf{G}\mathbf{G}^{H}\mathrm{diag}((\mathbf{h}_{s}\odot\mathbf{e})))^{T}\otimes(\mathrm{diag}(\mathbf{w}_{k}^{H}\mathbf{G}^{H})\mathbf{h}_{s}))}_{\boldsymbol{\Omega}_{1}}\boldsymbol{\theta}\mathbf{h}_{s}^{H}\mathrm{diag}(\mathbf{G}\mathbf{w}_{k})\boldsymbol{\theta}\nonumber\\
&=(\boldsymbol{\theta}\otimes\boldsymbol{\theta})^{H}\boldsymbol{\Omega}_{1}(\boldsymbol{\theta}\otimes\boldsymbol{\theta})\underbrace{\mathrm{vec}(\mathbf{h}_{s}^{H}\mathrm{diag}(\mathbf{G}\mathbf{w}_{k}))}_{\bar{\boldsymbol{\Omega}}_{1}}=(\boldsymbol{\theta}\otimes\boldsymbol{\theta})^{H}\boldsymbol{\Omega}_{1}(\boldsymbol{\theta}\otimes\boldsymbol{\theta})\bar{\boldsymbol{\Omega}}_{1}\nonumber\\
&=(\boldsymbol{\theta}\otimes\boldsymbol{\theta})^{H}\mathrm{vec}(\boldsymbol{\Omega}_{1}(\boldsymbol{\theta}\otimes\boldsymbol{\theta})\bar{\boldsymbol{\Omega}}_{1})=(\boldsymbol{\theta}\otimes\boldsymbol{\theta})^{H}\underbrace{(\bar{\boldsymbol{\Omega}}_{1}^{T}\otimes\boldsymbol{\Omega}_{1})}_{\tilde{\boldsymbol{\Omega}}_{1}}(\boldsymbol{\theta}\otimes\boldsymbol{\theta})=(\boldsymbol{\theta}\otimes\boldsymbol{\theta})^{H}\tilde{\boldsymbol{\Omega}}_{1}(\boldsymbol{\theta}\otimes\boldsymbol{\theta}),\nonumber\\
&g_{2}(\boldsymbol{\theta})=\circled{1}\times\circled{4}=(\boldsymbol{\theta}\otimes\boldsymbol{\theta})^{H}\tilde{\boldsymbol{\Omega}}_{2}(\boldsymbol{\theta}\otimes\boldsymbol{\theta}), g_{3}(\boldsymbol{\theta})=\circled{2}\times\circled{3}=
(\boldsymbol{\theta}\otimes\boldsymbol{\theta})^{H}\tilde{\boldsymbol{\Omega}}_{3}(\boldsymbol{\theta}\otimes\boldsymbol{\theta}),\nonumber\\
&g_{4}(\boldsymbol{\theta})=\circled{2}\times\circled{4}=(\boldsymbol{\theta}\otimes\boldsymbol{\theta})^{H}\tilde{\boldsymbol{\Omega}}_{4}(\boldsymbol{\theta}\otimes\boldsymbol{\theta}),\bar{g}_{2}(\boldsymbol{\theta})=\circled{5}\times\circled{3}=(\boldsymbol{\theta}\otimes\boldsymbol{\theta})^{H}\tilde{\boldsymbol{\Omega}}_{6}(\boldsymbol{\theta}\otimes\boldsymbol{\theta}),\nonumber\\
&\bar{g}_{3}(\boldsymbol{\theta})=\circled{5}\times\circled{4}=(\boldsymbol{\theta}\otimes\boldsymbol{\theta})^{H}\tilde{\boldsymbol{\Omega}}_{7}(\boldsymbol{\theta}\otimes\boldsymbol{\theta}),\bar{g}_{4}(\boldsymbol{\theta})=\circled{5}\times\circled{5}^{H}=(\boldsymbol{\theta}\otimes\boldsymbol{\theta})^{H}\tilde{\boldsymbol{\Omega}}_{8}(\boldsymbol{\theta}\otimes\boldsymbol{\theta}).\label{proA3__1}
\end{align}
\hrulefill
\end{figure*}
Based on the above operations in 
(\ref{proA3__1}), formulation (\ref{pro27}) holds. Similarly, by continuing to use the vectorization operation in \cite{b38}, 
$g_{5}(\boldsymbol{\theta})$, $g_{6}(\boldsymbol{\theta})$, $g_{7}(\boldsymbol{\theta})$, $g_{8}(\boldsymbol{\theta})$ can be written as 
(\ref{proA3_2}) at the top of the next page.

\begin{figure*}[htbp] 
\centering 
\vspace*{6pt} 
\begin{small}
\begin{align}
&g_{5}(\boldsymbol{\theta})=\circled{6}\times\circled{7}
=\sum_{d(j)\leq d(k)}(\boldsymbol{\theta}\otimes\boldsymbol{\theta})^{H}\underbrace{(\mathrm{diag}(\mathbf{h}_{u,i}^{H})\mathbf{G}\mathbf{w}_{k}\mathbf{w}_{j}^{H}\mathbf{G}^{H}\mathrm{diag}(\mathbf{h}_{u,\bar{k}}))^{T}\otimes(\mathrm{diag}(\rho_{k}\rho_{\bar{k}}\mathbf{w}_{k}^{H}\mathbf{G}^{H})\mathbf{h}_{u,i}))}_{\boldsymbol{\Xi}_{i,\bar{k},j}}\boldsymbol{\theta}\mathbf{h}_{u,\bar{k}}^{H}\mathrm{diag}(\mathbf{G}\mathbf{w}_{k})\boldsymbol{\theta}\nonumber\\
&=\sum_{d(j)\leq d(k)}(\boldsymbol{\theta}\otimes\boldsymbol{\theta})^{H}\boldsymbol{\Xi}_{i,\bar{k},j}(\boldsymbol{\theta}\otimes\boldsymbol{\theta})\underbrace{\mathrm{vec}(\mathbf{h}_{u,\bar{k}}^{H}\mathrm{diag}(\mathbf{G}\mathbf{w}_{k}))}_{\bar{\boldsymbol{\Xi}}_{i,\bar{k}}}=\sum_{d(j)\leq d(k)}(\boldsymbol{\theta}\otimes\boldsymbol{\theta})^{H}\boldsymbol{\Xi}_{i,\bar{k},j}(\boldsymbol{\theta}\otimes\boldsymbol{\theta})\bar{\boldsymbol{\Xi}}_{i,\bar{k}}\nonumber\\
&=\sum_{d(j)\leq d(k)}(\boldsymbol{\theta}\otimes\boldsymbol{\theta})^{H}\mathrm{vec}(\boldsymbol{\Xi}_{i,\bar{k},j}(\boldsymbol{\theta}\otimes\boldsymbol{\theta})\bar{\boldsymbol{\Xi}}_{i,\bar{k}})=(\boldsymbol{\theta}\otimes\boldsymbol{\theta})^{H}\underbrace{\sum_{d(j)\leq d(k)}(\bar{\boldsymbol{\Xi}}_{i,\bar{k}}^{T}\otimes\boldsymbol{\Xi}_{i,\bar{k},j})}_{\tilde{\boldsymbol{\Xi}}_{5}}(\boldsymbol{\theta}\otimes\boldsymbol{\theta})=(\boldsymbol{\theta}\otimes\boldsymbol{\theta})^{H}\tilde{\boldsymbol{\Xi}}_{5}(\boldsymbol{\theta}\otimes\boldsymbol{\theta}),\nonumber\\
&g_{6}(\boldsymbol{\theta})=\circled{7}\times\circled{8}=\sum_{d(j)\leq d(k)}(\boldsymbol{\theta}\otimes\boldsymbol{\theta})^{H}\underbrace{(\mathrm{diag}(\mathbf{h}_{u,\bar{k}}^{H})\mathbf{G}\mathbf{w}_{\bar{k}}\mathbf{w}_{j}^{H}\mathbf{G}^{H}\mathrm{diag}(\mathbf{h}_{u,i}))^{T}\otimes(\mathrm{diag}(\rho_{k}\rho_{k}\mathbf{w}_{\bar{k}}^{H}\mathbf{G}^{H})\mathbf{h}_{u,\bar{k}}))}_{\boldsymbol{\Omega}_{i,\bar{k},j}}\boldsymbol{\theta}\mathbf{h}_{u,\bar{k}}^{H}\mathrm{diag}(\mathbf{G}\mathbf{w}_{\bar{k}})\boldsymbol{\theta}\nonumber\\
&=\sum_{d(j)\leq d(k)}(\boldsymbol{\theta}\otimes\boldsymbol{\theta})^{H}\boldsymbol{\Omega}_{i,\bar{k},j}(\boldsymbol{\theta}\otimes\boldsymbol{\theta})\underbrace{\mathrm{vec}(\mathbf{h}_{u,i}^{H}\mathrm{diag}(\mathbf{G}\mathbf{w}_{\bar{k}}))}_{\bar{\boldsymbol{\Omega}}_{\bar{k},i}}=\sum_{d(j)\leq d(k)}(\boldsymbol{\theta}\otimes\boldsymbol{\theta})^{H}\boldsymbol{\Omega}_{i,\bar{k},j}(\boldsymbol{\theta}\otimes\boldsymbol{\theta})\bar{\boldsymbol{\Omega}}_{i,\bar{k}}\nonumber\\
&=\sum_{d(j)\leq d(k)}(\boldsymbol{\theta}\otimes\boldsymbol{\theta})^{H}\mathrm{vec}(\boldsymbol{\Omega}_{i,\bar{k},j}(\boldsymbol{\theta}\otimes\boldsymbol{\theta})\bar{\boldsymbol{\Omega}}_{i,\bar{k}})=(\boldsymbol{\theta}\otimes\boldsymbol{\theta})^{H}\underbrace{\sum_{d(j)\leq d(k)}(\bar{\boldsymbol{\Omega}}_{i,\bar{k}}^{T}\otimes\boldsymbol{\Omega}_{i,\bar{k},j})}_{\tilde{\boldsymbol{\Omega}}_{6}}(\boldsymbol{\theta}\otimes\boldsymbol{\theta})=(\boldsymbol{\theta}\otimes\boldsymbol{\theta})^{H}\tilde{\boldsymbol{\Omega}}_{6}(\boldsymbol{\theta}\otimes\boldsymbol{\theta}),\nonumber\\
&g_{7}({\boldsymbol{\theta}})=\circled{6}\times\circled{10}=\boldsymbol{\theta}^{H}\underbrace{\mathrm{diag}(\rho_{k}\mathbf{w}_{k}^{H}\mathbf{G}^{H})\mathbf{h}_{u,i}\mathbf{h}_{u,i}^{H}\mathrm{diag}(\mathbf{G}\mathbf{w}_{k}(\rho_{\bar{k}}\sigma_{\bar{k}}^{2}+\delta_{\bar{k}}^{2}))}_{\mathbf{D}_{i,\bar{k}}}\boldsymbol{\theta}=\boldsymbol{\theta}^{H}\mathbf{D}_{i,\bar{k}}\boldsymbol{\theta},\nonumber\\
&g_{8}({\boldsymbol{\theta}})=\circled{8}\times\circled{11}=\boldsymbol{\theta}^{H}\underbrace{\mathrm{diag}(\rho_{\bar{k}}\mathbf{w}_{k}^{H}\mathbf{G}^{H})\mathbf{h}_{u,\bar{k}}\mathbf{h}_{u,\bar{k}}^{H}\mathrm{diag}(\mathbf{G}\mathbf{w}_{k}(\rho_{k}\sigma_{k}^{2}+\delta_{k}^{2}))}_{\bar{\mathbf{D}}_{i,\bar{k}}}\boldsymbol{\theta}=\boldsymbol{\theta}^{H}\bar{\mathbf{D}}_{i,\bar{k}}\boldsymbol{\theta}.\label{proA3_2}
\end{align}
\end{small}
\hrulefill
\end{figure*}
Based on (\ref{proA3_2}), SIC constraint is rewritten as 
\begin{align}
&g_{5}(\boldsymbol{\theta})=f_{1},~g_{6}(\boldsymbol{\theta})=f_{2},~
f_{1}-f_{2}\leq g_{7}({\boldsymbol{\theta}})-g_{8}({\boldsymbol{\theta}})\nonumber\\
&=\boldsymbol{\theta}^{H}\tilde{\mathbf{D}}_{i,\bar{k}}\boldsymbol{\theta}, \label{proA3_3}
\end{align}
where $\tilde{\mathbf{D}}_{i,\bar{k}}=\bar{\mathbf{D}}_{i,\bar{k}}-\mathbf{D}_{i,\bar{k}}$. Based on the above considerations, (\ref{pro28}) and (\ref{pro29}) hold.

\section{BCD Algorithm's Convergence Proof}\label{appD}
In this section, we discuss the convergence of \textbf{Algorithm}~\ref{algo1}. The convergence of the BCD algorithm primarily depends on four stages, which we need to analyze in detail. We define $\mathbf{w}_{k}$, $\boldsymbol{\Theta}$, $d(k)$ and $\rho_{k}$ as the $t$-th iteration solution of problem (\ref{pro_15}), (\ref{pro25}), (\ref{pro41}), (\ref{pro46}). In $4$-th step of \textbf{Algorithm~\ref{algo1}}, the transmit beamforming vector $\mathbf{w}_{k}^{(t+1)}$ is obtained  for given $\boldsymbol{\Theta}^{(t)}$, $d(k)^{(t)}$ and $\rho_{k}^{(t)}$. Thus, we have
\begin{small}
\begin{align}
\mathcal{G}(\mathbf{w}_{k}^{(t+1)},\boldsymbol{\Theta}^{(t)},d(k)^{(t)},\rho_{k}^{(t)})\leq\mathcal{G}(\mathbf{w}_{k}^{(t)},\boldsymbol{\Theta}^{(t)},d(k)^{(t)},\rho_{k}^{(t)}).\label{proA4_14}
\end{align}
\end{small}%
In $5$-th step of \textbf{Algorithm~\ref{algo1}}, RIS phase shift $\boldsymbol{\Theta}^{(t+1)}$ is obtained  for given $\mathbf{w}_{k}^{(t+1)}$, $d(k)^{(t)}$ and $\rho_{k}^{(t)}$. Thus, we have 
\begin{small}
\begin{align}
&\mathcal{G}(\mathbf{w}_{k}^{(t+1)},\boldsymbol{\Theta}^{(t+1)},d(k)^{(t)},\rho_{k}^{(t)})\leq \mathcal{G}(\mathbf{w}_{k}^{(t+1)},\boldsymbol{\Theta}^{(t)},d(k)^{(t)},\nonumber\\
&\rho_{k}^{(t)}).\label{proA4_15}
\end{align}
\end{small}%
Similarly, in $6$-th step of \textbf{Algorithm~\ref{algo1}}, SIC order $d(k)^{(t+1)}$ is obtained  for given $\mathbf{w}_{k}^{(t+1)}$ is obtained  for given $\boldsymbol{\Theta}^{(t+1)}$, $\rho_{k}^{(t)}$. Thus, we have 
\begin{small}
\begin{align}
&\mathcal{G}(\mathbf{w}_{k}^{(t+1)},\boldsymbol{\Theta}^{(t+1)},d(k)^{(t+1)},\rho_{k}^{(t)})\geq \mathcal{G}(\mathbf{w}_{k}^{(t+1)},\boldsymbol{\Theta}^{(t+1)},d(k)^{(t)}\nonumber\\
&,\rho_{k}^{(t)}).\label{proA4_16}
\end{align}
\end{small}%
Finally, in $7$-th step of \textbf{Algorithm~\ref{algo1}}, PS ratio factor $\rho_{k}^{(t+1)}$ is obtained  for given $\mathbf{w}_{k}^{(t+1)}$,$\boldsymbol{\Theta}^{(t+1)}$,$d(k)^{(t+1)}$. Thus, we have 
\begin{small}
\begin{align}
&\mathcal{G}(\mathbf{w}_{k}^{(t+1)},\boldsymbol{\Theta}^{(t+1)},d(k)^{(t+1)},\rho_{k}^{(t+1)})\geq \mathcal{G}(\mathbf{w}_{k}^{(t+1)},\boldsymbol{\Theta}^{(t+1)},d(k)^{(t+1)}\nonumber\\
&,\rho_{k}^{(t)}).\label{proA4_17}
\end{align}
\end{small}%
Based on (\ref{proA4_14})-(\ref{proA4_17}), we have
\begin{small}
\begin{align}
&\mathcal{G}(\mathbf{w}_{k}^{(t+1)},\boldsymbol{\Theta}^{(t+1)},d(k)^{(t+1)},\rho_{k}^{(t+1)})\geq \mathcal{G}(\mathbf{w}_{k}^{(t)},\boldsymbol{\Theta}^{(t)},d(k)^{(t)}\nonumber\\
&,\rho_{k}^{(t)}).\label{proA4_18}
\end{align}
\end{small}%
This indicates that the objective function's value is both monotonic and non-increasing, ensuring the convergence of \textbf{Algorithm~\ref{algo1}} due to the finite upper bounds of the four subproblems.



\end{document}